\documentclass[useAMS,usenatbib]{mn2e}

\usepackage{amsmath, amssymb, amsfonts}    
\usepackage{journals}
\usepackage{graphicx}   
\usepackage{verbatim}   
\usepackage{color}      

\title[General relativistic MHD simulations of accretion around Sgr A*: How important are radiative losses?]{GRMHD simulations of accretion onto Sgr A*: How important are radiative losses?}
\author[S. Dibi et al.]{S. Dibi$^{1}$\thanks{E-mail:
s.dibi@uva.nl (Salom\'e D.)},  S. Drappeau$^{1}$, P. C. Fragile$^{2}$\thanks{KITP Visiting Scholar, Kavli Institute for Theoretical Physics, Santa Barbara, CA}, S. Markoff$^{1}$, and  J. Dexter$^{3}$ \\
$^{1}$Astronomical Institute ``Anton Pannekoek", University of Amsterdam, Postbus 94249,
1090 GE Amsterdam, The Netherlands\\
$^{2}$ Department of Physics and Astronomy, College of Charleston, 66 George St., Charleston SC 29424, United States\\
$^{3}$Theoretical Astrophysics Center and Department of Astronomy,
University of California, Berkeley, CA 94720-3411, United States}
\begin{document}

\date{in original form 2012 May 23}

\pagerange{\pageref{firstpage}--\pageref{lastpage}} \pubyear{2012}

\maketitle

\label{firstpage}

\begin{abstract}

  We present general relativistic magnetohydrodynamic (GRMHD)
  numerical simulations of the accretion flow around the supermassive
  black hole in the Galactic centre, Sagittarius A* (Sgr A*). The
  simulations include for the first time radiative cooling processes
  (synchrotron, bremsstrahlung, and inverse Compton) self-consistently
  in the dynamics, allowing us to test the common simplification of
  ignoring all cooling losses in the modeling of Sgr A*. We confirm
  that for Sgr A*, neglecting the cooling losses is a reasonable
  approximation if the Galactic centre is accreting
  below $\sim 10^{-8} M_{\odot}~{\rm yr^{-1}}$ i.e. $\dot{M} < 10^{-7}
  \dot{M} _{Edd}$. But above this limit, 
  we show that radiative losses should be taken into account as
  significant differences appear in the dynamics and the resulting
  spectra when comparing simulations with and without cooling. This
  limit implies that most nearby low-luminosity active galactic nuclei
  are in the regime where cooling should be taken into account.
	
  We further make a parameter study of axisymmetric gas accretion
  around the supermassive black hole at the Galactic centre. This
  approach allows us to investigate the physics of gas accretion in
  general, while confronting our results with the well studied and
  observed source, Sgr A*, as a test case.  We confirm that the nature
  of the accretion flow and outflow is strongly dependent on the
  initial geometry of the magnetic field.  For example, we find it
  difficult, even with very high spins, to generate powerful outflows
  from discs threaded with multiple, separate poloidal field loops.
\end{abstract}


\section{Introduction}

Super-massive black holes (SMBHs) of millions to billions of 
solar masses are believed to exist in the centre of most galaxies. The
Galactic center black hole candidate, Sgr
A*, is the closest and best studied SMBH, making it the perfect source
to test our understanding of galactic nuclei systems in general. This
compact object was first observed as a radio source by
\citet{balick74}.  Since then, observations have constrained important
parameters of Sgr A*, such as its mass and distance, estimated at
$M=4.3\pm0.5\times10^6M_\odot$ and $D=8.3\pm0.35~{\rm kpc}$,
respectively \citep{reid93, schodel02, ghez08, gillessen09}. The
accretion rate has also been constrained by polarisation measurements,
using Faraday rotation arguments \citep{aitken00, bower03, marrone07}
and is estimated to be in the range $2\times10^{-9}<\dot{M}
<2\times10^{-7} M_{\odot}~{\rm yr^{-1}}$. Other key
parameters, such as the spin, inclination, and magnetic field
configuration, are still under investigation.

Multi-wavelength observations of Sgr A* have been performed from the
radio to the gamma ray (see reviews by \citealt{ melia01, genzel10}, and
references therein). More recently, important progress has been
achieved in the infrared \citep[IR; e.g.,][]{schoedel11} and sub-millimeter
(sub-mm) domains. All observations agree that Sgr A* is a very 
under-luminous and weakly accreting black hole; indeed its accretion
rate is lower than has been observed in any other accreting system.

In the near future, the next milestone will be observing the first
black hole shadow from Sgr A* \citep{falcke00,dexter10} with the
proposed ``Event Horizon Telescope'' \citep{doeleman08, doeleman09, fish11} thanks to the capabilities of very long base interferometry at sub-mm wavelengths.
  Such a detection would be the first direct evidence for a black hole event
  horizon, and may also constrain the spin of Sgr A*. 
  In order to make accurate predictions for testing 
  with the Event Horizon Telescope, however, we need to have reliable models for the
  plasma conditions and geometry in the accretion (in/out)flow.
  General relativistic magnetohydrodynamic (MHD) simulations offer significant promise
  for this class of study, as they can provide both geometrical and
  spectral predictions.

  Sgr A* has already been modeled in several numerical studies 
  \citep[e.g.,][]{goldston05, moscibrodzka09,dexter09,dexter10, hilburn10,
    shcherbakov10,shiokawa12,dolence12,dexterfragile12}. All of these models consist of two separate codes:
  a GRMHD code describing the dynamics, and then a subsequent code to
  calculate the radiative emission based on the output of the first
  one. The under-luminous and under-accreting state of Sgr A* ($L_{\rm
    bol}\simeq 10^{-9} L_{\rm Edd}$ and $L_{\rm X-Ray}\simeq 10^{33}$ erg/s in the 0.5 to 10 keV band, where $L_{\rm Edd}$
  is the Eddington luminosity; \citealt{baganoff03}), is the common
  argument given to justify ignoring the cooling losses and simplify
  the description of Sgr A*. Even though this approach seems
  reasonable, especially for the peculiar case of Sgr A*, we propose
  to quantify to what extent it really applies.  This question is
  especially important if one wants to extend these studies to more
  typical nearby low-luminosity active galactic nuclei (LLAGN). An
  important example is the nuclear black hole in M87, which is the
  other major target for mm-VLBI and is significantly more luminous
  than Sgr A* ($L/L_{\rm Edd} \sim 10^{-6}-10^{-4}$).

To test whether or not ignoring the radiative cooling losses is a reasonable approximation, we need to take into account the radiative losses self-consistently in the dynamics, which is now possible with the \emph{Cosmos++} astrophysical fluid dynamics code \citep{anninos05,fragile09}.  By allowing the gas to cool, energy can be liberated from the accretion flow, potentially changing the dynamics of the system. The basic result that we show in this paper is that cooling is indeed negligible at the lowest end of the possible accretion rate range of Sgr A*, but plays an increasingly important role in the dynamics and the resulting spectra for higher accretion rates (relevant for most nearby LLAGN and even at the high end of the range of Sgr A*).

  This paper is organized as follows: In Section 2 we describe the new
  version of \emph{Cosmos++} used to perform this study. In Section 3
  we present the initial set-up of the simulations. In Section 4 we
  show the importance of the accretion rate parameter and assess the
  effect of including radiative cooling.  In Section 5 we perform a
  parameter survey to investigate the influence of the initial
  magnetic field configuration and the spin of the SMBH (we also
  discuss the effect of a retrograde spin on the dynamics) on the
  resulting accretion disc structure and outflow. In Section 6 we end
  with our conclusions and outlook. The spectra generated from these
  simulations are analyzed in detail in a companion paper (Drappeau et
  al., in preparation, hereafter referred to in the text as D12).

\section{Numerical Method}
 
 \subsection{GRMHD Equations}
 
Within \textit{Cosmos++} we solve the following set of conservative, general relativistic MHD equations, including radiative cooling,
\begin{eqnarray}
 \partial_t D + \partial_i (DV^i) &=& 0 ~,  \label{eqn:de} \\
 \partial_t {\cal E} + \partial_i (-\sqrt{-g}~T^i_0) &=&
      -\sqrt{-g}~T^\kappa_\lambda~\Gamma^\lambda_{0\kappa} + \sqrt{-g}~\Lambda u_0 ~,
    \label{eqn:en} \\
 \partial_t {\cal S}_j + \partial_i (\sqrt{-g}~T^i_j) &=&
      \sqrt{-g}~T^\kappa_\lambda~\Gamma^\lambda_{j\kappa} - \sqrt{-g}~\Lambda u_j ~,
    \label{eqn:mom} \\
 \partial_t \mathcal{B}^j + \partial_i (\mathcal{B}^j V^i - \mathcal{B}^i V^j) &=&
     0 ~, 
      \label{eqn:ind} 
\end{eqnarray}
where $D=W\rho$ is the generalized fluid density, $W=\sqrt{-g} u^0$ is the generalized boost factor, $V^i=u^i/u^0$ is the transport velocity, $u^\mu = g^{\mu \nu} u_\nu$ is the fluid 4-velocity, $g_{\mu\nu}$ is the 4-metric, $g$ is the 4-metric determinant, 
\begin{eqnarray}
{\cal E} & = & -\sqrt{-g} T^0_0 \nonumber \\
 & = & -W u_0 \left(\rho h + 2 P_B \right) - \sqrt{-g}\left(P + P_B\right) + \sqrt{-g} b^0 b_0
\end{eqnarray}
is the total energy density, $h = 1 + \epsilon + P/\rho$ is the specific enthalpy, $\epsilon$ is the specific internal energy, $P$ is the fluid pressure, $P_B$ is the magnetic pressure, $T^\kappa_\lambda$ is the stress-energy tensor, $\Lambda$ is the cooling function, and
\begin{equation}
{\cal S}_\mu = \sqrt{-g} T^0_j = W u_j \left(\rho h + 2 P_B \right) - \sqrt{-g} b^0 b_j
\end{equation}
is the covariant momentum density. With indices, $\Gamma$ indicates the geometric connection coefficients of the metric.  Without indices, $\Gamma$ is the adiabatic index.  For this work we use an ideal gas equation of state (EOS),
\begin{equation}
P = (\Gamma-1)\rho\epsilon ~,
\end{equation}
with $\Gamma = 5/3$.

There are multiple representations of the magnetic field in our
equations: $b^\mu$ is the magnetic field measured by an observer
comoving with the fluid, which can be defined in terms of the dual of
the Faraday tensor $b^\mu \equiv u_\nu {^*F^{\mu\nu}}$, and
$\mathcal{B}^j = \sqrt{-g} B^j$ is the boosted magnetic field
3-vector. The un-boosted magnetic field 3-vector $B^i = {^*F^{\mu i}}$
is related to the comoving field by
\begin{equation}
B^i = u^0 b^i - u^i b^0 ~.
\end{equation}
The magnetic pressure is $P_B = b^2/2 = b^\mu b_\mu/2$. Note that, unlike in previous versions of \emph{Cosmos++}, we have absorbed the factor of $\sqrt{4\pi}$ into the definition of the magnetic fields, so-called Lorentz-Heaviside units ($B_{\rm LH}=B_{\rm cgs}/\sqrt{4\pi}$).

\subsection{GRMHD Solver}

We note that the MHD equations as written are all in the form of conservation equations
\begin{equation}
\partial_t \mathbf{U}(\mathbf{P}) + \partial_i \mathbf{F}^i(\mathbf{P}) = \mathbf{S}(\mathbf{P}) ~, \label{eqn:cons}
\end{equation}
where $\mathbf{U}$, $\mathbf{F}^i$, and $\mathbf{S}$ represent the
conserved quantities, fluxes, and source terms, respectively. These
are solved using a new High Resolution Shock Capturing (HRSC) scheme
recently added to the \emph{Cosmos++} computational astrophysics code
\citep{anninos05}.  The new HRSC scheme is modeled after the original
non-oscillatory central difference (NOCD) scheme of \emph{Cosmos}
\citep{anninos03}.  It also has many of the same elements as the
publicly available \emph{HARM} code \citep{gammie03,noble06}, although
with staggered magnetic fields (instead of HARM's zone-centered fields) and the inclusion of
a self-consistent, \emph{physical} cooling model.  More details of the
new scheme are provided in a separate paper \citep{fragile12}.

Briefly, as in other, similar conservative codes, the fluxes,
$\mathbf{F}^i$, are determined using the Harten-Lax-Van Leer (HLL) 
approximate Riemann solver \citep{HLL83}
\begin{equation}
\mathbf{F} = \frac{c_{min} \mathbf{F}_R + c_{max} \mathbf{F}_L - c_{max} c_{min} ( \mathbf{U}_R - \mathbf{U}_L )}{c_{max} + c_{min}} ~.
\end{equation}
A slope-limited parabolic extrapolation gives $\mathbf{P}_R$ and
$\mathbf{P}_L$, the primitive variables at the right- and left-hand
side of each zone interface.  From $\mathbf{P}_R$ and $\mathbf{P}_L$,
we calculate the right- and left-hand conserved quantities
($\mathbf{U}_R$ and $\mathbf{U}_L$), the fluxes $\mathbf{F}_R =
\mathbf{F}(\mathbf{P}_R)$ and $\mathbf{F}_L =
\mathbf{F}(\mathbf{P}_L)$, and the maximum right- and left-going waves
speeds, $c_{\pm , R}$ and $c_{\pm , L}$. The bounding wave speeds are
then $c_\mathrm{max} \equiv \mathrm{max}(0, ~c_{+,R}, ~c_{+,L})$ and
$c_\mathrm{min} \equiv -\mathrm{min}(0, ~c_{-,R}, ~c_{-,L})$.

For stability, the equations are integrated using a staggered leapfrog
method (2nd order in time). First a half-time step, $\Delta t/2$, is
taken to project the conserved variables $\mathbf{U}^n$ forward to
$n+1/2$. From these, a new set of primitives $\mathbf{P}^{n+1/2}$ can
be computed. These intermediate primitives are then used in
calculating the fluxes and source terms needed for the full time step,
$\Delta t$.  To find the new primitive variables, $\mathbf{P}$, after
each update cycle, we use either the $2D$ or $1D_W$ numerical methods
of \citet{noble06}.

Along with the evolution equations, we must also satisfy the
divergence-free constraint $\partial_j \mathcal{B}^j = 0$.  To
accomplish this, \emph{Cosmos++} now has the option to use a staggered
magnetic field with a Constrained-Transport (CT) update scheme akin to the
Newtonian version described in \citet{stone09}.

\subsection{Cooling}

In the present work, we assume the whole system is optically thin, i.e. radiation escapes freely from the system. However, for the calculation of the radiation we consider the appropriate optical depth of the gas at a given location and time, which depends on the state. This approximation takes into account the optical depth without performing radiative transfer, and is valid as long as the (assumed thermal) peak of the radiating particle distribution corresponds to energies greater than the self-absorption frequency, which is almost always the case for the regions under study.

The radiative cooling term, $\Lambda$, in equations (\ref{eqn:en}) and
(\ref{eqn:mom}), is the cooling function introduced to \emph{Cosmos++}
in \citet{fragile09} and is based on \citet{esin96}, which includes
treatments of bremsstrahlung, synchrotron, and the inverse-Compton
enhancement of each of these.  Since Sgr A* is optically thin, we do not need to worry about multi-scattering events in the treatment of the inverse Compton emission. The cooling function is calculated
locally in each zone from the fluid density, electron temperature,
magnetic field strength, and scale height [$\Lambda = \Lambda(\rho,
T_e, b^2, H)$]. The local electron temperature is recovered by
assuming a constant ion-to-electron temperature ratio, $T_i/T_e$,
(fixed for each simulation) and calculating the ion temperature from
the fluid variables $T_i = \mu m_H P/k_B \rho$, where $\mu = 1.69$ is
the mean molecular weight (appropriate for Solar abundances), $m_H$ is the mass of hydrogen, and $k_B$ is
Boltzman's constant.  The temperature scale height is defined as
$H\equiv T_e^4/\vert\nabla (T_e^4)\vert$. Our modifications to the
original \citet{esin96} cooling function are described in
\citet{fragile09} and are also described in more detail in our companion paper (D12).

In taking the ion and electron temperatures to be held in a fixed
ratio, we are assuming there is some efficient coupling process at
work between the ions and electrons.  The same was done in
\citet{esin96} and \citet{fragile09}, except here we relax the
assumption somewhat so that the coupling does not have to be exact
($T_i/T_e$ can be greater than 1).  This procedure is not entirely
self-consistent since the expression for ion-electron collisions in
bremsstrahlung cooling [Equation (17) of \citet{fragile09}] assumes
the ions and electrons have the same temperature.  Therefore, whenever
$T_i > T_e$, we are actually underestimating the amount of
bremsstrahlung cooling.  However, because bremsstrahlung is such a
minor contributor to the cooling, this omission is not significant in
the context of our current work.  Obviously, assuming a fixed ratio of $T_i/T_e$
is not likely to be physically accurate.  However, it is the current
standard in most simulations \citep[e.g.][]{moscibrodzka09,dexter10}
and recent comparisons with more sophisticated treatments has so far
not shown large discrepancies (Dexter, Quataert, priv. comm.).

Because the cooling time of the gas, $t_\mathrm{cool} = \rho
\epsilon/\Lambda$, can sometimes be shorter than the MHD timestep,
$\Delta t_\mathrm{MHD} \sim \Delta x/V$, where $\Delta x$ and $V$ are
the characteristic zone length and velocity, respectively, we allow
the cooling routine to operate on its own shorter timestep (subcycle),
if necessary.  To prevent runaway loops inside the cooling function,
we limit the cooling subcycle to 4 steps, with a maximum change to the
specific internal energy each subcycle of $\Delta \epsilon = 0.5
\epsilon$.  Even with these restrictions, the cooling has the
potential to decrease the internal energy (and hence temperature) by
up to nearly 95\% each MHD cycle.

\subsection{Spectra}

Our method for generating spectra is described in detail in our companion paper (D12).  However, we want to mention one important point related to how we present spectra in this paper.  MHD simulations of accretion discs, such as the ones used in our work, generically show significant variability due to the stochastic nature of MRI-generated turbulence and magnetic
reconnection.  Together, these effects can sometimes lead to large flares, similar
to those observed in the solar corona.  Since our goal is
to depict ``representative'' spectra, we have to make a choice about
what to consider ``typical'' behavior.  One option would be to produce many simulations using the same initial setup, but with different random seeds, and then select only those simulations that give desirable results.  This was the procedure in \citet{moscibrodzka09}. 

We have instead chosen a different approach, in that we perform only one simulation for each set of parameters, and then present the median spectrum for each simulation as the representative one.  By choosing the median value of all spectra, we minimize the impact of ``extreme'' episodes, especially a few very brief synchrotron/inverse Compton flares that appear to be attributable to numerical limitations of the simulations, most particularly the forced axisymmetry.  For error bars, we give the ``1-sigma'' variation about the median, i.e. the limits within which 68\% of the spectra fall. For example, if we have 50 individual spectra (corresponding to 50 different simulation times), as is typically the case in this work, then for each spectral energy bin, we drop the 8 highest and 8 lowest data points.  In fact dropping just the 4 highest and lowest data points already gives similar results, but we use the 1-sigma values throughout.  

In this paper we only include spectra computed over the inner $15 r_G$ of the simulation domain.  We have confirmed, via comparison with spectra computed using the entire simulation domain, that most of the emission comes from this inner region.  The emission from the jets in our simulations is completely
negligible because of their extremely low density.  This extreme level
of magnetic domination is likely partially a byproduct of adhering to
ideal MHD, where matter cannot effectively load the jets.  We hope to
explore the jet contribution to the spectrum more in future works.

\section{Simulation Setup}

\subsection{Initial State}

We start each simulation with a torus of gas around a compact object
situated at the origin.  The mass of the central compact object is set
to the mass of Sgr A* ($M_{\rm BH}=4.3\times 10^{6} M_{\odot}$), and
the initial density profile inside the torus is chosen to produce a desired mass accretion rate, somewhere in the range $10^{-9}$ and $10^{-7}
M_{\odot}~{\rm yr^{-1}}$ (depending on the simulation) at the inner
grid boundary.

The free parameters that describe the torus are the black hole spin
$a_*=a/M_\mathrm{BH} = c J/G M_{\rm BH}^2$ (where $J$ is the angular
momentum of the black hole and $r_G = GM_{\rm BH}/c^2 \simeq
6.35\times 10^{11} {\rm cm} \simeq 2.06\times 10^{-7} {\rm pc}$ ), the
inner radius of the torus ($r_{\rm in}=15 r_G$), the radius of the
pressure maximum of the torus ($r_{\rm center}=25 r_G$), and the
power-law exponent ($q=1.68$) used in defining the specific angular
momentum distribution,
\begin{equation}
\ell = -u_\phi/u_t \propto \left( -\frac{g_{t \phi}+\ell g_{tt}}{\ell g_{\phi
\phi} + \ell^2 g_{t \phi}}\right)^{q/2-1} ~,
\end{equation}
We then follow the procedure in \citet{chakrabarti85} to solve for the initial internal energy distribution of the torus $\epsilon(r,\theta)$. This sets its initial temperature profile, $T_0=(\Gamma-1)(\mu
m_H/k_B)\epsilon$. For the purpose of initialization, we assume an isentropic equation of state
$P=\rho \epsilon(\Gamma-1)=\kappa \rho^\Gamma$, so that now the density is given by $\rho = \left[
\epsilon(\Gamma-1)/\kappa \right]^{1/(\Gamma-1)}$. We can then use the parameter $\kappa$ to set the density (and mass) normalization of the initial torus. 

\begin{figure}
 \includegraphics[scale=0.15]{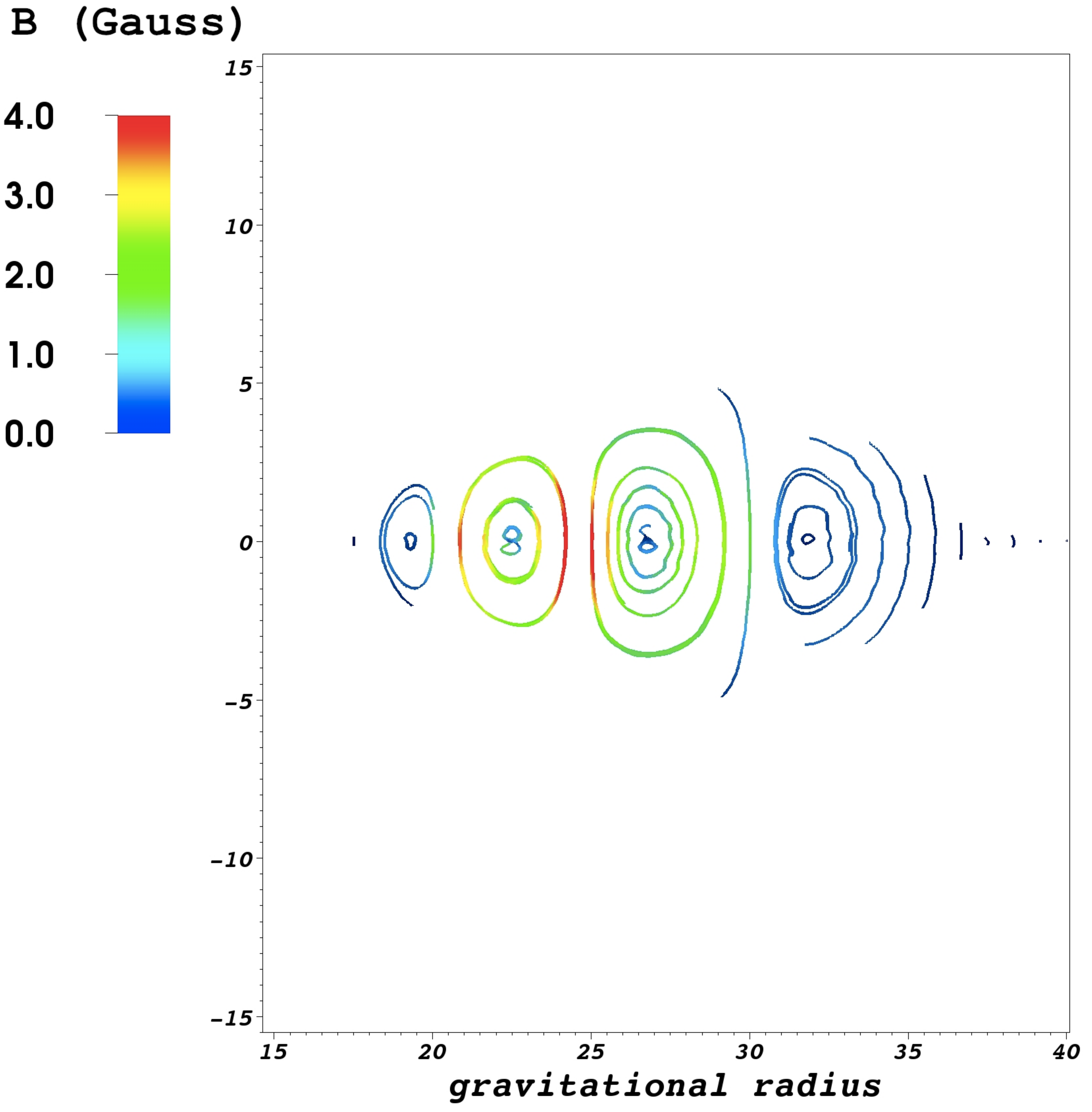}
 \includegraphics[scale=0.15]{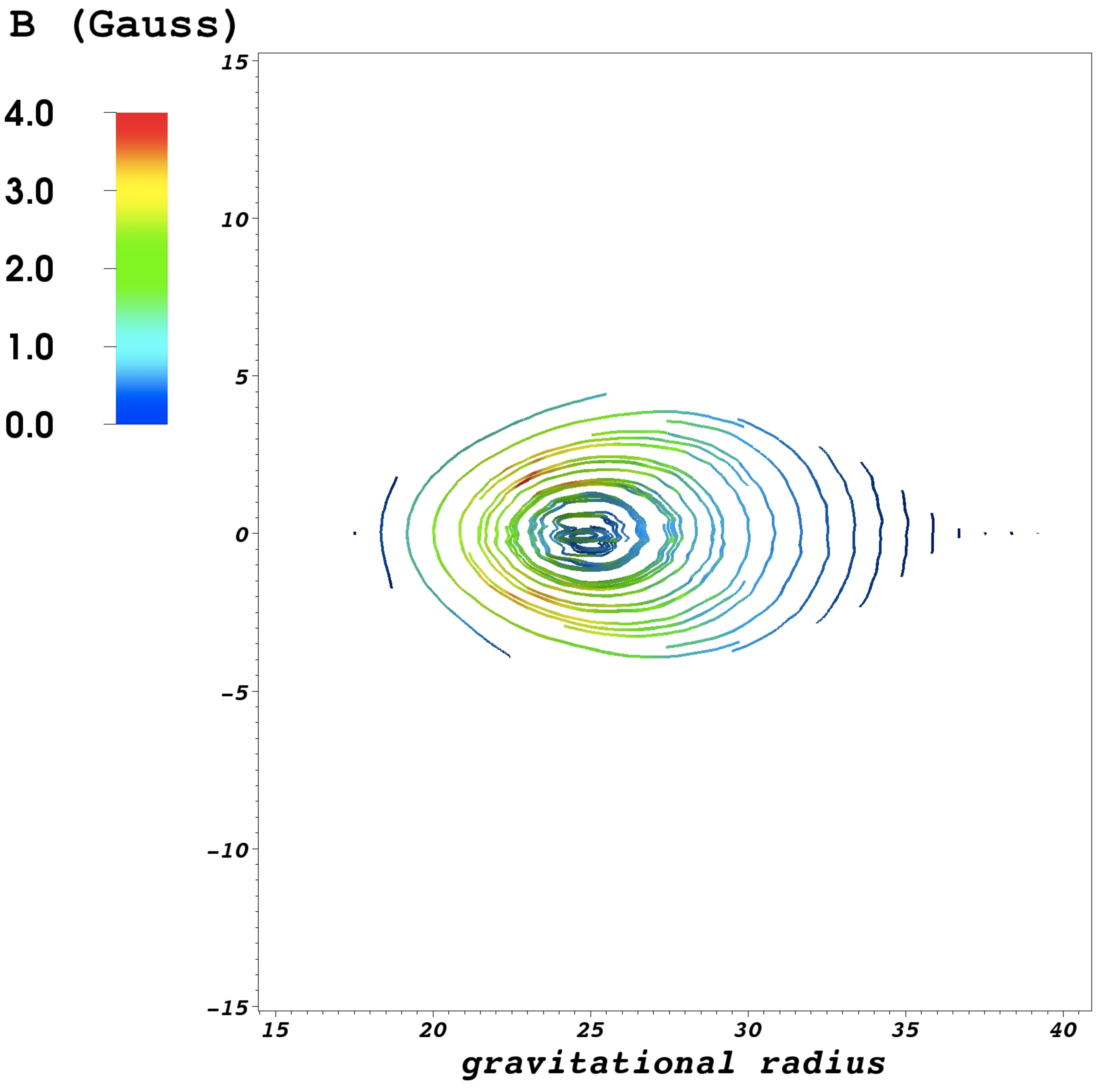}
\caption{\textit{Initial magnetic field configuration for the four poloidal loop case (top panel) and the single poloidal loop case (bottom panel). The colour scales with the value of the magnetic field.}} 
\label{Initial_MagneticField}
\end{figure}

The torus is seeded with weak poloidal magnetic field loops to drive the magnetorotational instability (MRI) \citep{balbus91}.  The non-zero spatial components are $\mathcal{B}^r = -\partial_\theta A_\phi$ and $\mathcal{B}^\theta = \partial_r A_\phi$, where
\begin{equation}
A_\phi = \left\{ \begin{array}{ccc}
          C(\rho-\rho_{\rm cut})^2 \sin \left[ 4 N \log (r/S) \right] & \mathrm{for} & \rho\ge\rho_{\rm cut}~, \\
          0                  & \mathrm{for} & \rho<\rho_{\rm cut}~,
         \end{array} \right.
\label{eq:torusb}
\end{equation}
$N$ is the number of field loop centers, and $S = 1.1 r_{\rm in}$.  In
this work we consider configurations with $N = 1$ and 4, as
illustrated in Figure \ref{Initial_MagneticField}, in order to
investigate the effect of changing $N$. The parameter $\rho_{\rm
  cut}=0.5 \rho_{\rm max,0}$ is used to keep the field a suitable
distance inside the surface of the torus initially, where $\rho_{\rm
  max,0}$ is the initial density maximum within the torus.  Using the
constant $C$, the field is normalized such that initially $\beta_{\rm
  mag} =P/P_B \ge \beta_{\rm mag,0}=10$ throughout the torus, where
$P_B$ is the magnetic pressure. This normalization is to ensure that
the initial magnetic field is weak.  This way of initializing the
field is slightly different than what other groups have done,
e.g. \citet{beckwith08}, who used a volume integrated $\beta_{\rm
  mag}$, or \citet{mckinney09}, who used $\beta_{\rm mag} = P_{\rm
  avg}/P_{B,{\rm avg}}$, to set the field strength \citep[see][for a
discussion of the different methods]{mckinney12}.  We also tested one
simulation with $\beta_{\rm mag,0}=50$. The higher $\beta$ case ends
up with a density (over $r < 15 r_G$) that is about a factor of 2
higher, a temperature that is 30-50\% colder, a scale-height that is
about 40\% thinner, and a magnetic pressure that is about 50\%
smaller.  The higher $\beta$ case contributes $\sim$70\% more flux in
the infrared waveband, and $\sim$86\% more in the X-ray band, compared
to the weaker case (see D12).

The fact that $\beta_{mag,0}$ affects our final result is not surprising since our simulations do not reach a saturated (magnetically arrested or magnetically choked) state \citep{igumenshchev03,igumenshchev08,tchekhovskoy11,mckinney12}.  Our resulting field strength is dictated by how much magnetic field is made available to the black hole through our initial conditions. 

In the background region not specified by the torus solution, we set
up a low density non-magnetic gas. Numerical floors are placed on
density and energy density with the following forms: $\rho_{\rm floor} =
10^{-4} \rho_{\rm max,0} r^{-1.5}$ and $e_{\rm floor}=\rho\epsilon =
10^{-6} \rho_{\rm max,0} r^{-2.5}$. These floors are never applied
within the disc, nor in most of the background region. They are only
seldom applied very close to the outer boundary (the few last external
cells), and more frequently along the vertical axis in the funnel
region.  The most commonly applied floor is that on the ratio $(\rho + \rho\epsilon)/P_B$.  Whenever this quantity drops below 0.01 (almost always within the jets), both $\rho$ and $\rho\epsilon$ are rescaled by a factor appropriate to maintain this ratio.  

\subsection{Grid}

All of the simulations are performed in 2.5 spatial dimensions (all
three spatial components of vector quantities are evolved in a single
azimuthal slice, although symmetry is assumed in the azimuthal
direction) using a spherical polar coordinate grid. The grid used in
the majority of the simulations consists of 256 radial zones and 256
zones in polar angle.

In the radial direction, we use a logarithmic coordinate of the form
$\eta \equiv 1.0 + \ln (r/r_{\rm BH})$, where $r_\mathrm{BH} = r_G (1+\sqrt{1-a_*})$ is
the radius of the black hole horizon; therefore, $\Delta r /r =
\mathrm{constant}$. The inner and outer radial
boundaries are set at $0.9 r_\mathrm{BH}$ and $120 r_G$,
respectively. Note that, because we use the Kerr-Schild form of the
Kerr metric, we are able to place the inner radial boundary some
number of zones (usually 4) inside the black hole horizon. In
principle, this choice should keep the inner boundary causally disconnected
from the flow, since MHD signals cannot physically radiate out from
within the event horizon.  This situation is preferable to having the inner
boundary of the grid outside the event horizon, which can lead to
artificial behavior within the flow.  The spatial resolution near the
black hole horizon is $\Delta r \approx 0.023 r_G$; near the initial
pressure maximum of the torus, it is $\Delta r \approx 0.45 r_G$.  We
use outflow boundary conditions at both the inner and outer boundaries
(the MHD primitive variables are copied from interior zones into ghost
zones, except the radial velocity component, which is zeroed out if it
would lead to inflow onto the grid).

Since we consider a fairly small radial range, there is some concern that our jets (discussed in Section 5.1) might be affected by reflections of waves off the outer boundary of the grid.  To test this, we performed one simulation, B4S9T3M9Ce, that used an extended radial grid with $r_\mathrm{max} = 1.1 \times 10^4 r_G$.  We found that most properties of this simulation were very similar to the corresponding results done on the smaller (default) grid.  The deviations were consistent with those expected from using a different random perturbation seed, as was the case here since the new simulation used a different processor distribution which affects the random seeding.

\begin{figure}
 \includegraphics[scale=0.13]{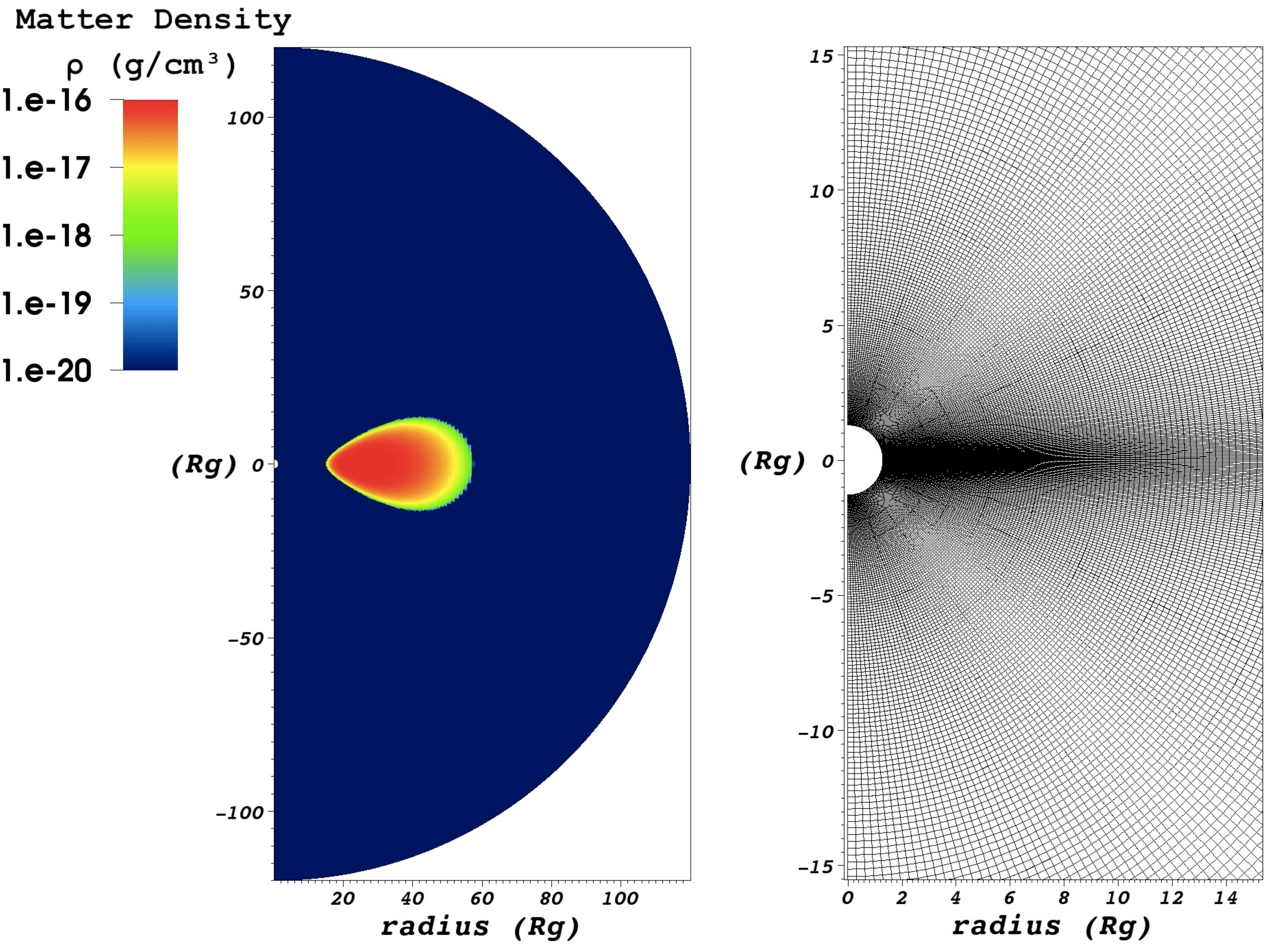}
\caption{\textit{Initial setup of the simulation. The colour/saturation scales with the density. The number of cells scales inversely with distance from the black hole and angle away from the equatorial plane. (The left panel shows the full computational domain while the right panel only shows the region of interest closer to the central SMBH.)}} 
\label{InitialConfiguration}
\end{figure}

In the angular direction, we include the full range $0 \le \theta \le
\pi$, with reflecting boundary conditions applied at the poles
(meaning that the MHD primitive variables are copied from interior
zones into ghost zones, with the sign reversed on the $\theta$
component of all vector quantities, plus the staggered magnetic field
component zeroed out at the pole).  We use a concentrated latitude
coordinate $x_2$ of the form $\theta = x_2 + \frac{1}{2} (1 - h) \sin
(2 x_2)$ with $h = 0.3$, which gives us a better resolution near the
midplane ($r_{\rm center} \Delta \theta = 0.1 r_G$).  Figure
\ref{InitialConfiguration} shows the initial state of the simulation
together with the grid resolution.

Our choice of resolution is sufficient to ensure that the fastest
growing poloidal field MRI modes are well resolved
[$\lambda_\mathrm{MRI} \equiv 2\pi v_{\mathrm{A}_z}/\Omega \gtrsim 10
\Delta z$ \citep{hawley11}] for at least the first few orbits of the
simulation.  We also confirm that $\alpha_{\rm mag} = -B^r
B^\phi/P_B \approx 0.3$, as expected \citep{hawley11}.  We also
performed select simulations at one-half and at double our default
resolution to directly test the numerical convergence of our results
(simulations B4S9l and B4S9h in Table 1). We find very little
variation between our default resolution (such as in simulation B4S9)
and its higher resolution counterpart (B4S9h), suggesting our results are well
converged.  

\begin{figure}
 \includegraphics[scale=0.34]{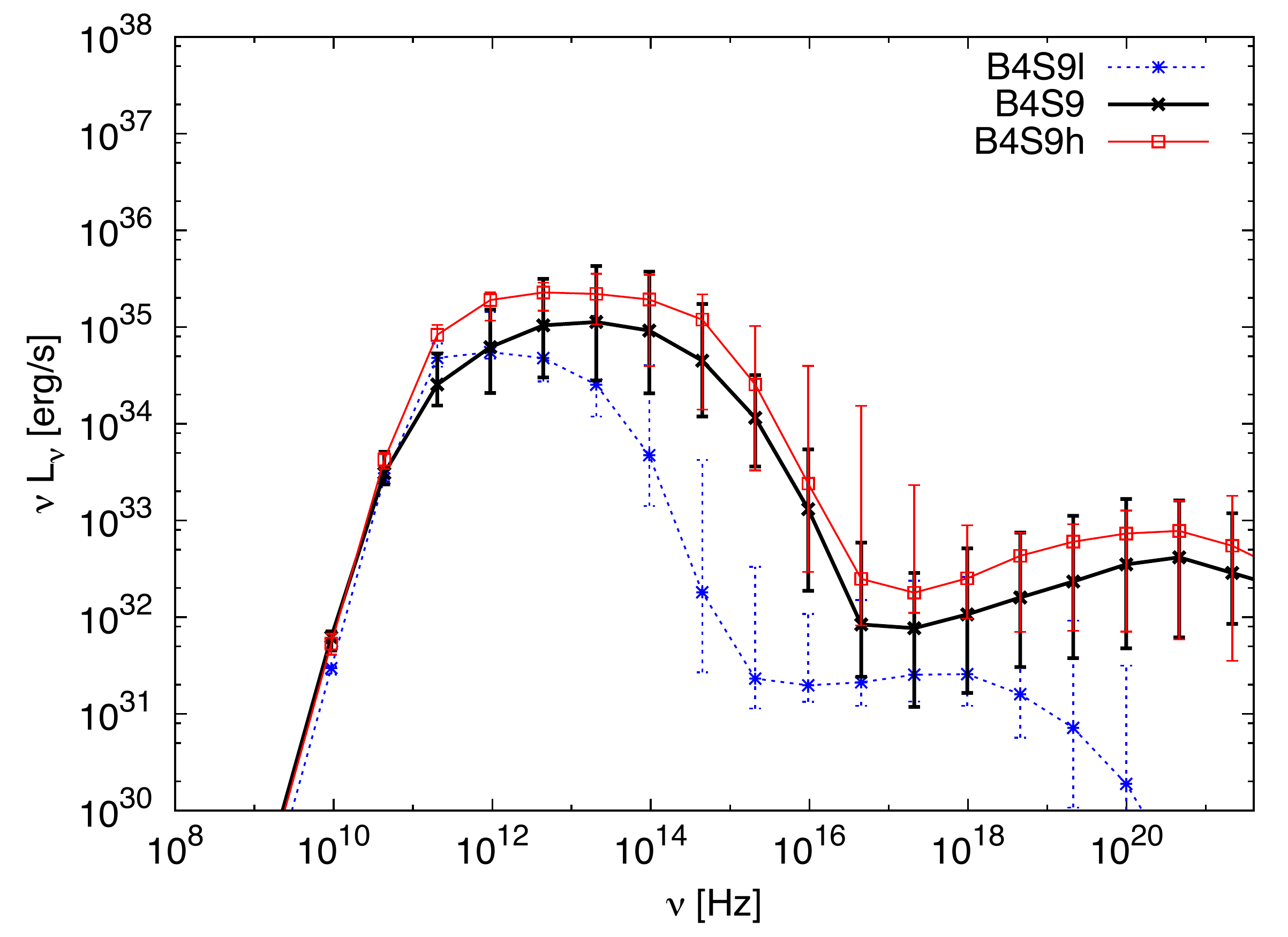}
\caption{\textit{Median broadband spectra taken from all timesteps in
    the interval 2.5-3.5 orbits for 3 simulations done at different resolutions. The
    error bars represent the 1 sigma limits about the median (see text for more details). The spectra contain two components: the synchrotron component responsible for the first bump, and the inverse Compton component at higher frequencies.}}
\label{ResolutionTest}
\end{figure}

Since our ultimate goal is to compare spectra from our simulations
with actual data, our true criterion for demonstrating reasonable
convergence is to compare spectra generated from simulations at
different resolutions.  Figure \ref{ResolutionTest} demonstrates that simulation B4S9, with
a resolution of $256\times256$, gives a spectrum that agrees with simulation B4S9h, with a resolution of $384\times384$, to within our error bars. On
the other hand, simulation B4S9l, with a resolution of $192\times128$, produces a
markedly different spectrum that diverges from the other two.
Recently, \citet{shiokawa12} confirmed that once a certain resolution
is achieved, further increases in resolution have little effect on the spectra.

\subsection{Time interval}
\label{sec:time_interval}

Because the simulations are performed assuming axial symmetry, the
anti-dynamo theorem stating that an axisymmetric magnetic field cannot be maintained via dynamo action \citep{Cowling33}, prevents
the MRI from being self-sustaining.  After an initial phase of
vigorous growth of the MRI channel modes, the turbulence gradually
dies out as the simulations progress.  This 2D approximation is a main caveat of the simulations we have run, and it is therefore important for us to carefully choose the time interval that we
wish to analyse.  We want it to be after the initial phase of MRI
growth, but before the mass accretion has begun to decay too
dramatically (longer simulation times are not necessarily beneficial in 2D for this reason).  Since we have a target mass accretion rate in mind for
each simulation, we can use that as a guide for selecting our time
interval.  

\begin{figure}
 \includegraphics[scale=0.35]{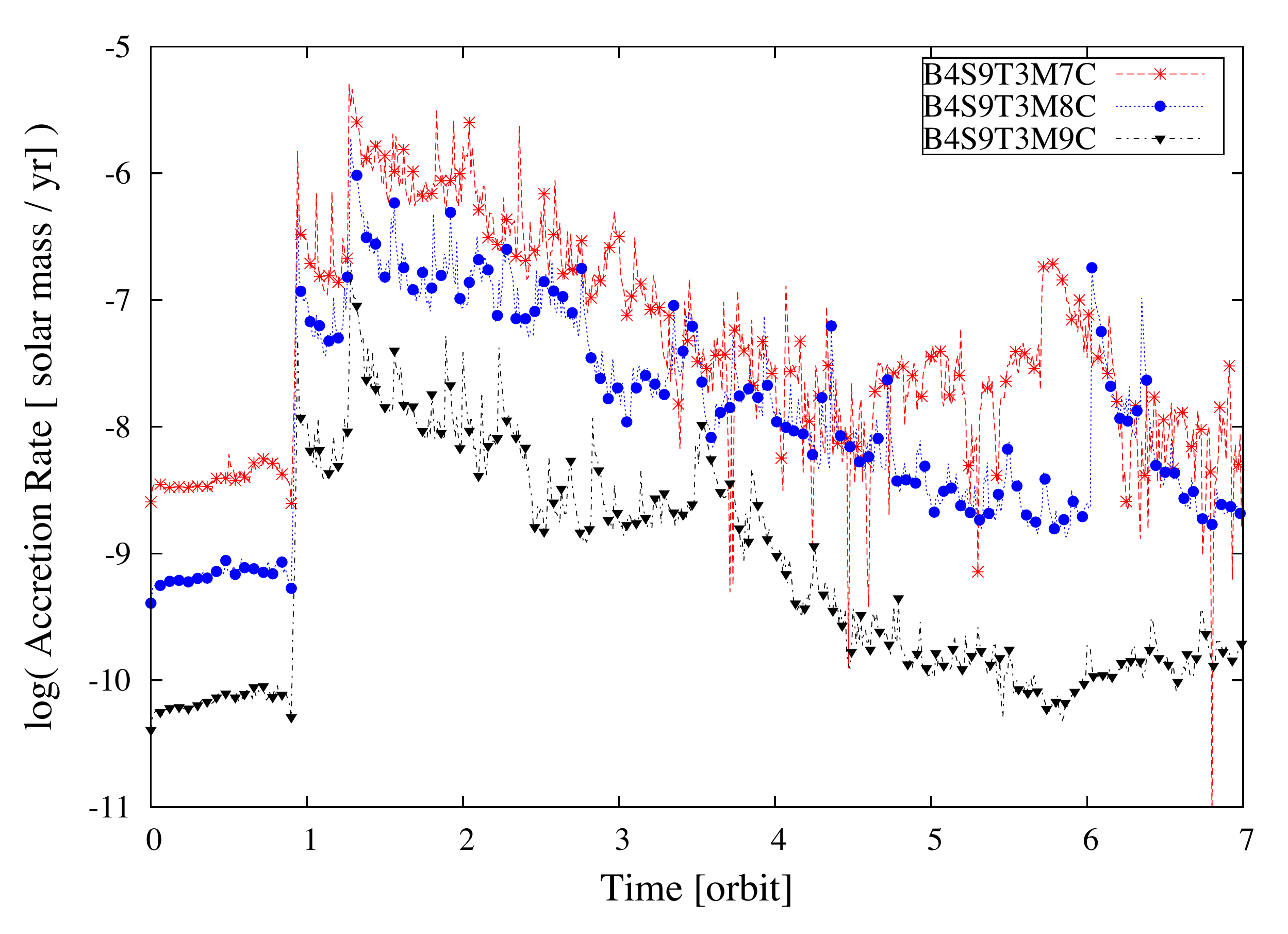}
\caption{\textit{Mass accretion rate at the event horizon of simulations B4S9T3M9C, B4S9T3M8C, and B4S9T3M7C as a function of time.}} 
\label{Mdot}
\end{figure} 

Figure \ref{Mdot} shows the variation of the accretion rate
as a function of time for three simulations that include cooling
(B4S9T3M9C, B4S9T3M8C, and B4S9T3M7C).  We see that it takes roughly
1.5 orbits to have the accretion reach its peak value, and that the
accretion dies out (returns to the background rate) after about orbit
5, where we are referring to the circular orbital period at $r=r_{\rm center}$, i.e. $t_{\rm orb} = 2\pi(g_{t\phi}\ell - g_{\phi\phi})/(g_{tt}\ell - g_{t\phi}) = 1.67\times10^4~{\rm s}$.  For these three simulations, the target accretion rates were
$10^{-9}$, $10^{-8}$, and $6\times 10^{-8} M_{\odot}~{\rm
  yr^{-1}}$, respectively. Figure \ref{Mdot} then suggests that an appropriate
interval would be between 2.5 and 3.5 orbits.  We therefore use this
as the standard time interval for analysis. in the rest of this paper,
$2.5t_{\rm orb} = t_{\rm min} \le t \le t_{\rm max} = 3.5t_{\rm orb}$.  We confirmed that our region of interest ($r \leq 15 r_G$) has reached inflow equilibrium over this time interval, based, for instance, on the mass flux being constant as a function of radius over this region.  Therefore, we can also be confident that our selected time interval is late enough not to be affected by transient solutions.

\section{Results: The importance of including cooling losses}

As we have stated, our primary goal in this paper is to assess the
importance of including radiative cooling losses self-consistently in
numerical simulations of LLAGN like Sgr A*.  We do this by comparing
two types of simulations: those that include radiative cooling losses
self-consistently in the dynamics (indicated by a ``C'' in the
simulation name or cooling ``ON'' in Table 1) and those that neglect
them (no ``C'', cooling ``OFF'').  We performed a set of 25 different
simulations to assess the importance of radiative cooling losses, as
well as study the influence of parameters such as the initial magnetic
field configuration, the spin, the ion to electron temperature ratio,
and the mass accretion rate.

Table 1 summarizes all of the simulations performed. The name given to
each simulation is derived from its parameter choices in the order
they appear in the table.  We also include the average and root mean
square (rms) fluctuations of the mass accretion rates actually
measured in our simulations.  This value is to be compared with our
target values of $\dot{M}$.  In simulations without cooling, this
value is arbitrary since the initial density (or mass) of the disc is
a free parameter (other than the requirement that the total mass of
the disc $M_{\rm disc}$ be negligible compared to $M_{\rm BH}$ since
we ignore the self-gravity of the disc).  Its choice does not affect
the subsequent evolution of the simulation, so can be freely rescaled
after the fact.  This freedom does not extend to the simulations that
include cooling, since the cooling function, $\Lambda$, depends
directly on $\rho$.  Therefore, to test different mass accretion
rates, it is sufficient in simulations without cooling to do a single
simulation and then simply rescale it to each of the target accretion
rates, whereas in the simulations with cooling, a new simulation must
be conducted for each new mass accretion rate.  This physical scaling
is an important distinction between our work and that of previous
authors -- each of our simulations with cooling has a real, {\em
  physical} mass accretion rate associated with it; it is no longer
a parameter that can be freely fit to the data.  When comparing
simulations with cooling against those without, we always assume the
same initial accretion rate.

\begin{table*}
 \centering
 \begin{minipage}{140mm}
  \caption{\textit{Description of simulation parameters}}
  \begin{tabular}{@{}lccccccccc@{}}
  \hline
   Simulation  &  $B$ loops ($N$)    &  Spin ($a_*$)  &  $T_i/T_e$  &  Target $\dot{M}$  &  statistical av.  $\dot{M}$ &  Cooling  &  Resolution\\   
 \hline
B4S9T3M9C & 4 & 0.9 & 3 & $10^{-9}$ & 2.36$\pm 1.54 \times 10^{-9}$  & ON & $256 \times 256$ \\
B4S9T3M9Ct\footnote{Disk scale height of $H/r\approx 0.07$ instead of 0.12 as for B4S9T3M9C} & 4 & 0.9 & 3 & $10^{-9}$ & $3.05\pm2.22 \times 10^{-9}$ & ON & $256 \times 256$ \\
B4S9T3M9Cb\footnote{$\beta_\mathrm{mag,0}=50$ instead of 10} & 4 & 0.9 & 3 & $10^{-9}$ & $1.95\pm2.62 \times 10^{-9}$ & ON & $256 \times 256$ \\
B4S9T3M9Ce\footnote{$r_\mathrm{max} = 1.1\times10^4 r_G$ instead of $120 r_G$} & 4 & 0.9 & 3 & $10^{-9}$ & $3.00\pm2.00 \times 10^{-9}$ & ON & $512 \times 256$ \\
B4S9 & 4 & 0.9 & - & - & - & OFF & $256 \times 256$ \\
B1S9T3M9C & 1 & 0.9 & 3 & $10^{-9}$ & 7.93$\pm 9.27 \times 10^{-9}$ & ON & $256 \times 256$ \\
B1S9 & 1 & 0.9 & - & - & - & OFF & $256 \times 256$ \\
B4S0T3M9C & 4 & 0 & 3 & $10^{-9}$ &  2.88$ \pm 2.02 \times 10^{-9}$ & ON & $256 \times 256$ \\
B4S5T3M9C & 4 & 0.5 & 3 & $10^{-9}$ & 5.33$\pm 4.87 \times 10^{-9}$ & ON & $256 \times 256$ \\
B4S7T3M9C  & 4 & 0.7 & 3 & $10^{-9}$ & 5.38$\pm 4.06 \times 10^{-9}$ & ON & $256 \times 256$ \\
B4S98T3M9C & 4 & 0.98 & 3 & $10^{-9}$  & 3.98$\pm 3.36 \times 10^{-9}$ & ON & $256 \times 256$ \\
B4S9rT3M9C & 4 & -0.9 & 3 & $10^{-9}$ & 0.64$\pm 0.47 \times 10^{-9}$  & ON & $256 \times 256$ \\
B4S9T1M9C & 4 & 0.9 & 1 & $10^{-9}$ &  3.85$\pm 3.50 \times 10^{-9}$ & ON & $256 \times 256$ \\
B4S9T10M9C & 4 & 0.9 & 10 & $10^{-9}$ & 3.62$ \pm 3.90 \times 10^{-9}$ & ON & $256 \times 256$ \\
B4S9l & 4 & 0.9 & - & - & - & OFF & $192 \times 128$ \\
B4S9h & 4 & 0.9 & - &  - & - & OFF & $384 \times 384$ \\
B4S9T3M8C & 4 & 0.9 & 3 & $10^{-8}$ &  4.99 $\pm3.80 \times 10^{-8}$ &  ON & $256 \times 256$ \\
B4S9T3M7C  & 4 & 0.9 & 3 & $6.3\times 10^{-8}$ & 1.69 $\pm 1.39 \times 10^{-7}$ & ON & $256 \times 256$ \\
B4S0T3M7C & 4 & 0 & 3 & $6.3\times 10^{-8}$ & 2.16 $\pm 1.46 \times 10^{-7}$ & ON & $256 \times 256$ \\
B4S5T3M7C  & 4 & 0.5 & 3 & $6.3\times 10^{-8}$ & 2.03$ \pm 1.85 \times 10^{-7}$  & ON & $256 \times 256$ \\
B4S75T3M7C  & 4 & 0.75 & 3 & $6.3\times 10^{-8}$ & 1.65$ \pm 1.15 \times 10^{-7}$ & ON & $256 \times 256$ \\
B4S98T3M7C  & 4 & 0.98 & 3 & $6.3\times 10^{-8}$ & 1.37$\pm 1.08 \times 10^{-7}$  & ON & $256 \times 256$ \\
B4S9rT3M7C  & 4 & -0.9 & 3 & $6.3\times 10^{-8}$ &  0.39$\pm 0.41 \times 10^{-7}$ & ON & $256 \times 256$ \\
B4S9T1M7C  & 4 & 0.9 & 1 & $6.3\times 10^{-8}$ & 1.53$\pm 1.02 \times 10^{-7}$ & ON & $256 \times 256$ \\
B4S9T10M7C  & 4 & 0.9 & 10 & $6.3\times 10^{-8}$ & 1.43$ \pm 0.88 \times 10^{-7}$ & ON & $256 \times 256$ \\
\hline
\end{tabular}
\end{minipage}
\end{table*}

\subsection{Effects of the cooling losses on the dynamics}

\begin{figure}
\includegraphics[scale=0.35]{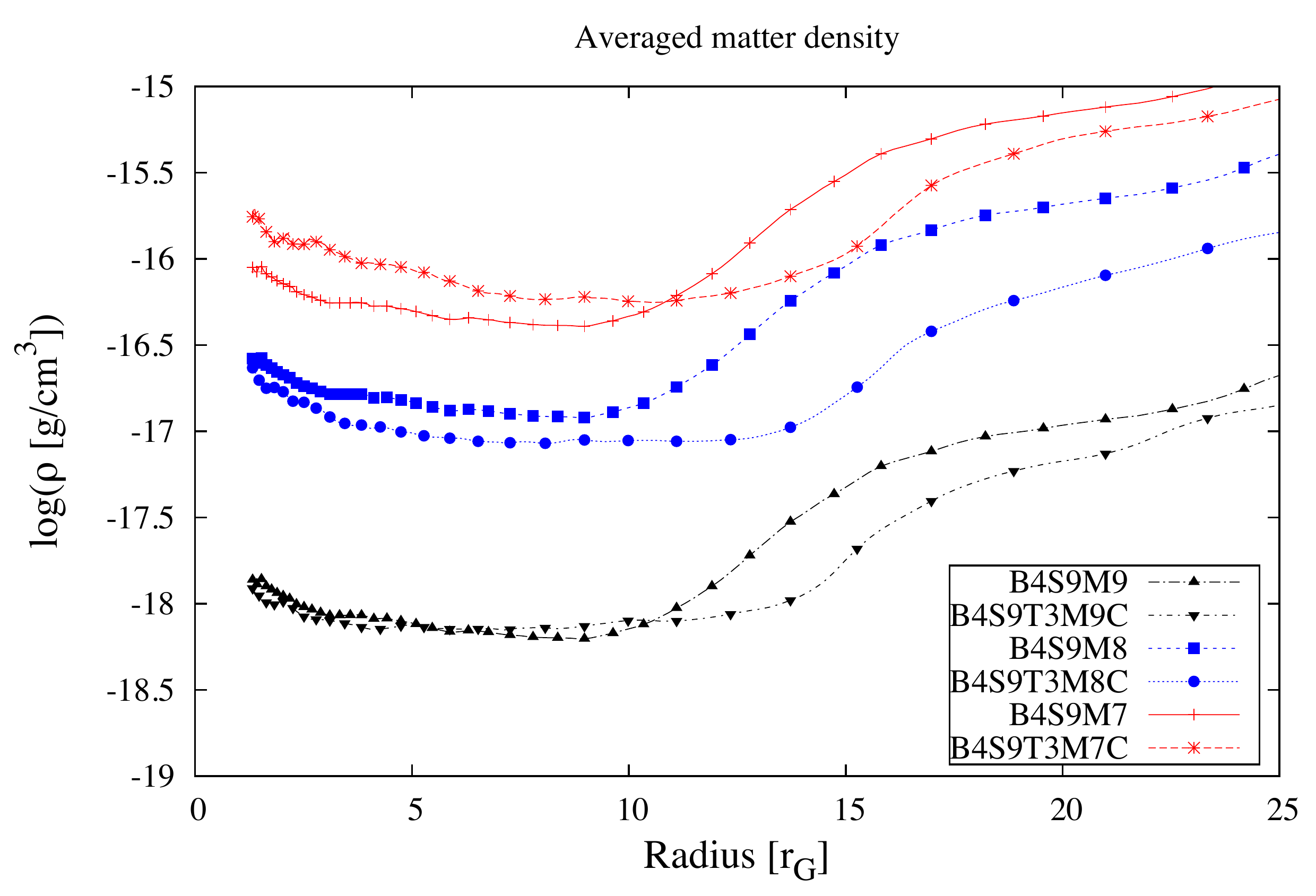}
\caption{\textit{This figure illustrates the relative importance of
    radiative losses on disc density. The plot shows the averaged
    density for simulations B4S9T3M9C, B4S9T3M8C, B4S9T3M7C and B4S9
    (rescaled to three different accretion rates). A time average is
    taken over the interval $t_{\rm min}-t_{\rm max}$, and at each
    radius we take an average over the shell.}} 
\label{Rho}
\end{figure}

\begin{figure}
\includegraphics[scale=0.135]{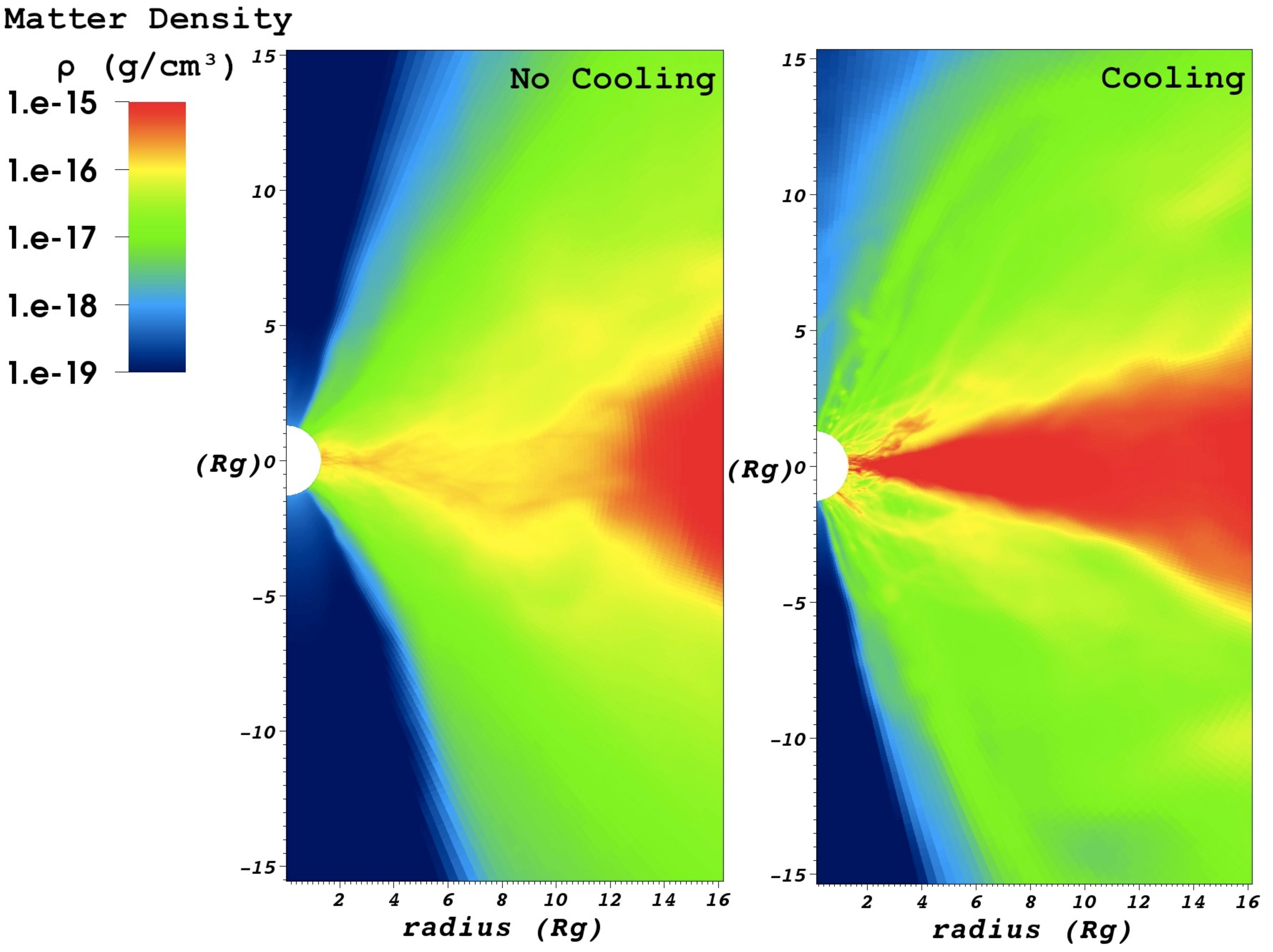}
\caption{\textit{Time-averaged matter density (in g/cm$^3$) of the system for simulations B4S9 and B4S9T1M7C, respectively. The simulations are initially identical; differences in the images are due entirely to cooling losses, included in the simulation in the right panel. The time average is taken over the interval $t_{\rm min}-t_{\rm max}$.}} 
\label{Rho_Cool}
\end{figure}

\begin{figure}
\includegraphics[scale=0.35]{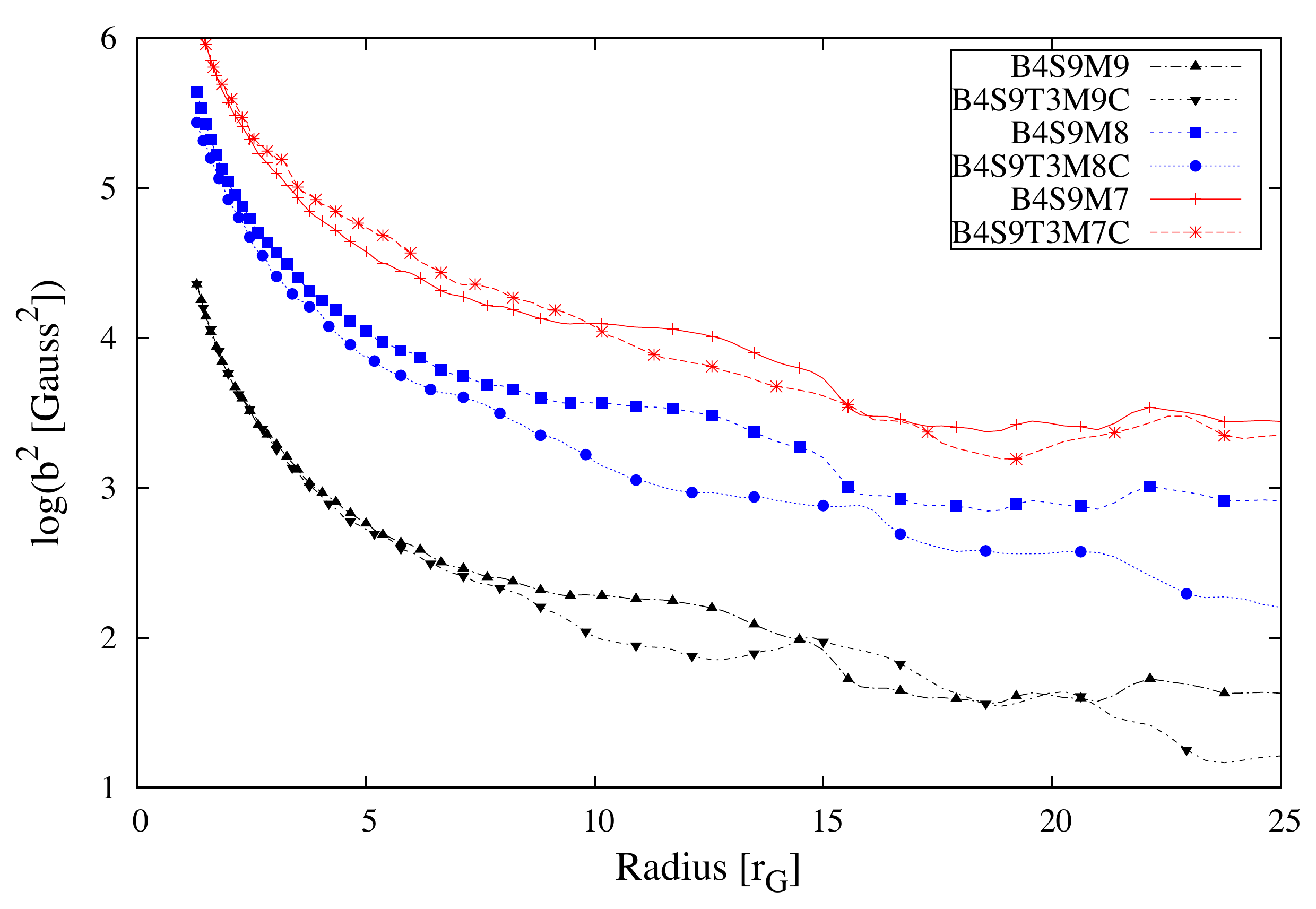}
\caption{\textit{Plot illustrating the relative importance of radiative losses on the resulting magnetic field. The plot shows the average of $P_B$ for simulations B4S9T3M9C, B4S9T3M8C, B4S9T3M7C and B4S9 (rescaled to three different accretion rates). A time average is taken over the interval $t_{\rm min}-t_{\rm max}$, and at each radius we take a density-weighted average over the shell.}} 
\label{B}
\end{figure}

\begin{figure}
\includegraphics[scale=0.35]{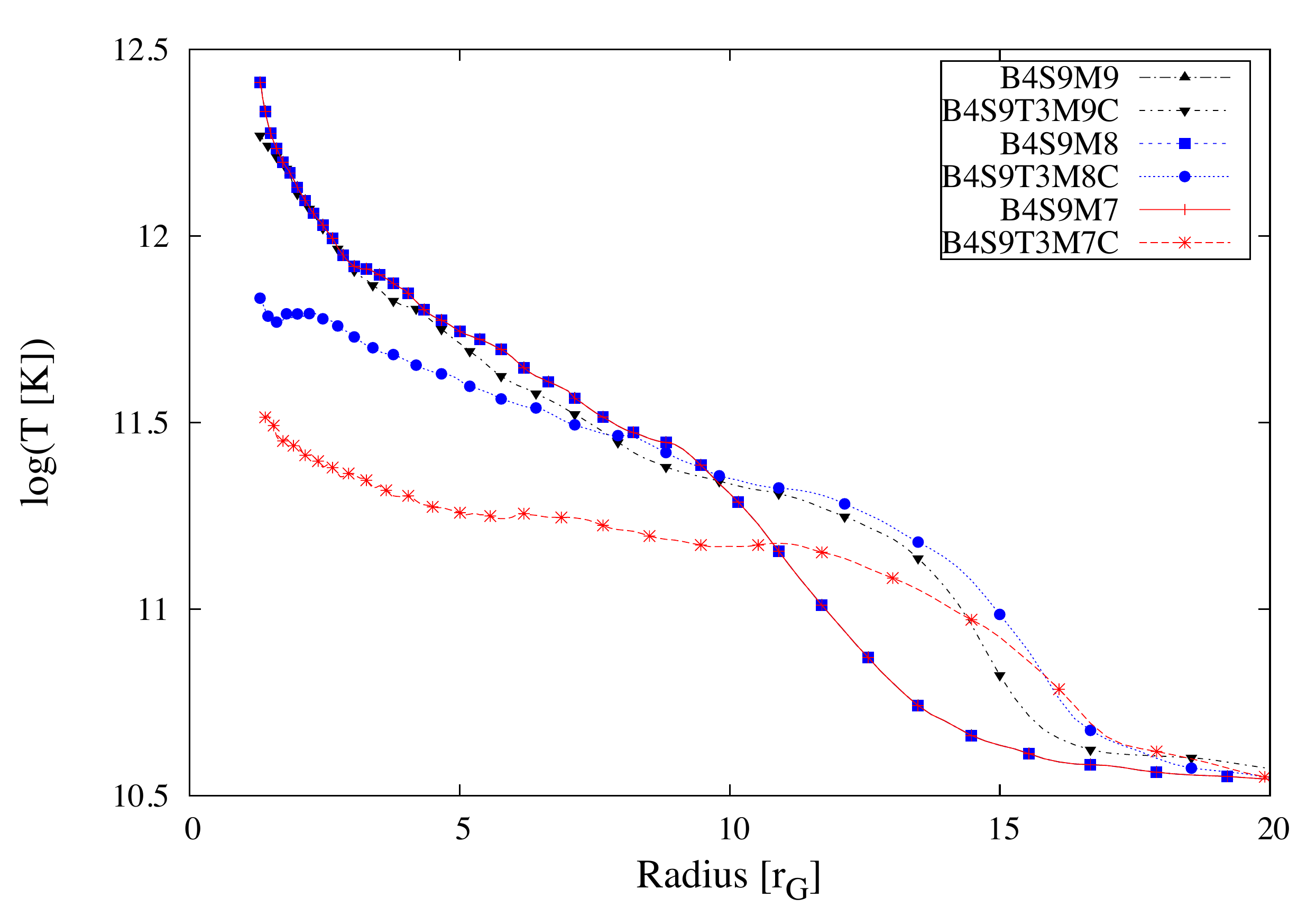}
\caption{\textit{Plot illustrating the relative importance of radiative losses on disc temperature. The plot shows the averaged temperature for simulations B4S9T3M9C, B4S9T3M8C, B4S9T3M7C and B4S9 (rescaled to three different accretion rates). A time average is taken over the interval $t_{\rm min}-t_{\rm max}$, and at each radius we take a density-weighted average over the shell. Because the temperature does not change when the density is rescaled, the three non-cooling curves (pink, blue, and orange) are exactly on top of each other whether $\dot{M}=10^{-9}, 10^{-8}, or \ 10^{-7} M_{\odot}~{\rm yr^{-1}}$}} 
\label{Temp}
\end{figure}

Since our cooling function $\Lambda$ depends on $\rho$, $T_e$, and
$b^2$ (as well as $H$), one way to assess the importance of cooling in
the simulations is to observe how these quantities compare between
simulations that include cooling and those that do not.  Differences
would indicate that the simulations with cooling are adjusting
dynamically to the radiative losses.  We present these comparisons in
Figures \ref{Rho} -- \ref{Temp}. 

Figure \ref{Rho} shows the time- and shell-averaged density, $\rho$,
for four different simulations: B4S9T3M9C, B4S9T3M8C, B4S9T3M7C, and
B4S9.  The simulation B4S9 has been scaled to three different accretion rates to be consistent with the three simulations that include cooling.  A
clear trend is apparent in this figure: as the mass accretion rate
increases, the differences between simulations with cooling and those
without also increases.  At our lowest accretion rate ($\dot{M}
\simeq 10^{-9}M_{\odot}~{\rm yr^{-1}}\simeq 10^{-8}\dot{M}_{\rm Edd}$), the differences are
relatively small ($\lesssim 30$\%), while at our highest accretion
rate ($\dot{M} \approx 2\pi \times 10^{-8}M_{\odot}~{\rm yr^{-1}}\simeq 6.6\times 10^{-7}\dot{M}_{\rm Edd}$),
they become more substantial ($\sim 50-70$\%).  Note that the apparent
``bump'' in density at $r \approx 15 r_G$ is simply indicating that
the initial mass reservoir (torus) has not fully redistributed itself
by this time in the simulation (as $r_{\rm in} = 15 r_G$).  The
enhanced density associated with the cooling simulations can also be
seen in Figure \ref{Rho_Cool}, which compares simulations B4S9 and
B4S9T3M7C.  With the cooling losses included, we end up with a
thinner, denser disc.

In Figure \ref{B} we show a similar plot for the density-weighted magnetic field magnitude, $b^2$.  Again the trend is that the differences between simulations with and without cooling become more pronounced at larger accretion rates even-though the trend is less obvious. We have a perfect match for cooling vs non cooling at the lowest accretion rate for the inner part of the disc, while the curves are distinct for the higher accretion rates. Also, the overall normalization of the magnetic field increases with mass accretion rate, which makes sense since the magnetic field is, roughly speaking, normalized by the mass density of the fluid.  Further, since we are assuming ideal MHD, the magnetic field is trapped in the fluid, so as the torus spreads into a disc, the magnetic flux will be carried with it.  Therefore, it is not surprising that the differences in Figure \ref{B} are not as large as in Figure \ref{Rho} for $\rho$.

\begin{figure}
\includegraphics[scale=0.135]{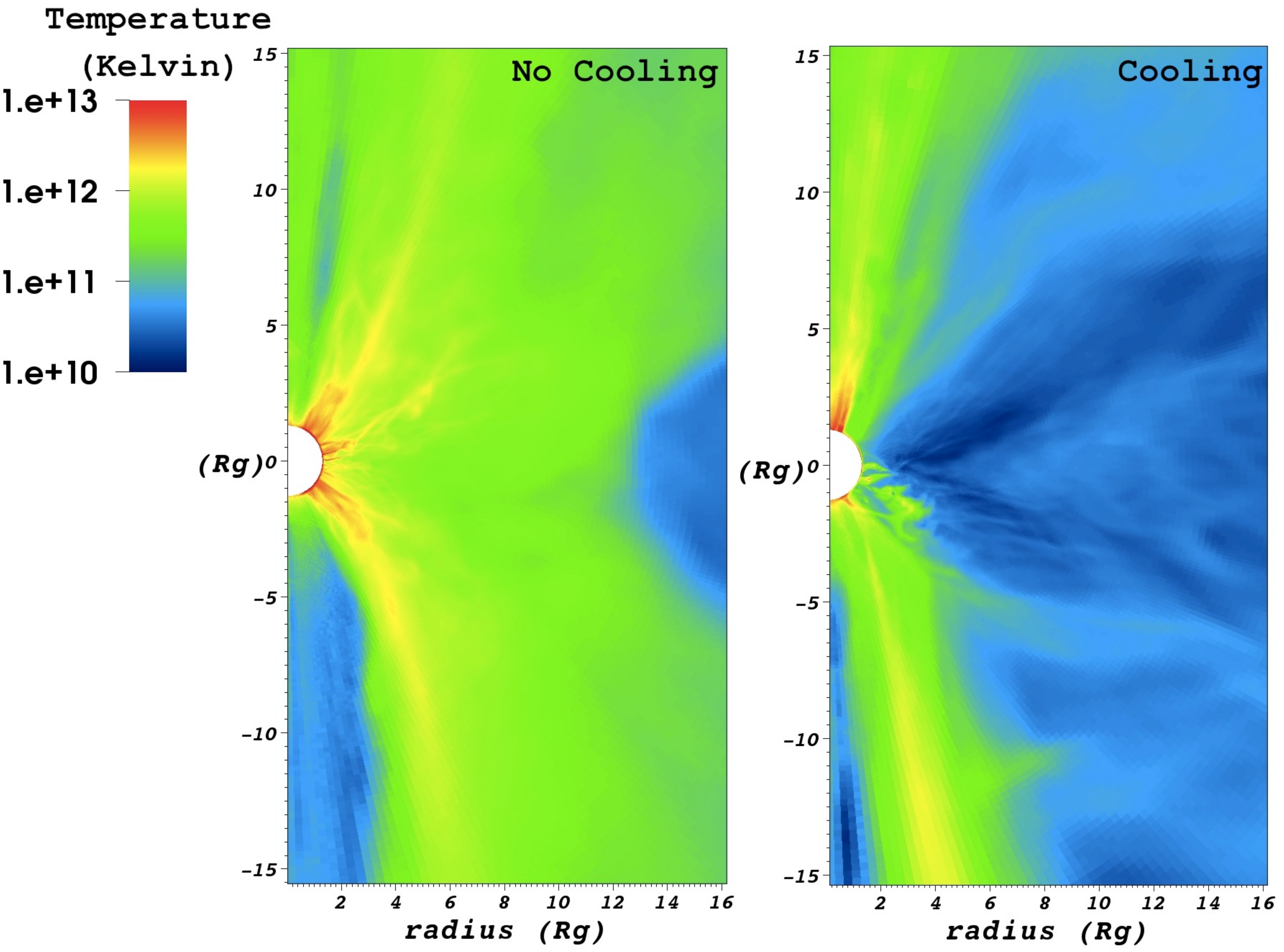}
\caption{\textit{Time-averaged temperature (in Kelvin) of the system for simulations B4S9 and B4S9T1M7C, respectively. The simulations are initially identical; differences in the images are due entirely to cooling losses, included in the simulation in the right panel. The time average is taken over the interval $t_{\rm min}-t_{\rm max}$. Note that the system is very dynamical, with many different time scales playing themselves out.  Even averaging over an orbital period does not give a perfectly symmetric system as analytic models would predict.}} 
\label{Temp_Cool}
\end{figure}

The density-weighted temperature (representative of the disc temperature) is shown in Figure \ref{Temp}, which helps illustrate one of our main conclusions in this paper.  The close agreement between the non-cooling simulations and simulation B4S9T3M9C (target $\dot{M} \approx 10^{-9}M_{\odot}~{\rm yr^{-1}}$) suggests that at this level of accretion, radiative losses are not important.  However, for $\dot{M} \gtrsim 10^{-8}M_{\odot}~{\rm yr^{-1}}$ i.e. $\dot{M} \gtrsim 10^{-7} \dot{M}_{Edd}$, the differences in temperatures of the discs dramatically increases, with the highest $\dot{M}$ simulation almost an order of magnitude colder than the lowest, at small radii.  This result suggests that only for the very lowest luminosity AGN (Sgr A* being the only known candidate example at this time) is it reasonable to neglect radiative cooling losses in numerical simulations.  Figure \ref{Temp_Cool} demonstrates the dramatic difference in temperature, up to 2 order of magnitude in the region of interest (the inner disc), between our highest accretion rate simulation, B4S9T3M7C, and a simulation that neglected cooling (scaled to the same accretion rate).  Note, especially, that only with cooling does any gas with $T < 3\times 10^{11}~\mathrm{K}$ reach the black hole event horizon. When cooling is turned on, the lost radiation pressure results in a thinner disk, and this in turn compresses the gas and magnetic fields, so we end up with a thinner but denser disk  in the inner part of the disk, while the  temperature is much cooler due to the radiative losses.

\subsection{Effects of the cooling losses on the resulting spectra}

Using numerical simulations together with spectral modeling codes to
fit Sgr A* has already been considered by many authors
\citep[e.g.][]{goldston05, moscibrodzka09, dexter10, hilburn10,
  shcherbakov10}.  What is new about our work is that it is the first that a numerical study of Sgr A* includes the effects of radiative cooling self-consistently.

In our companion paper (D12), we present the
details of producing broadband spectra from our simulations, together
with a parameter study that we use to constrain the physical
environment of Sgr A*.  In this paper we focus on how including
cooling affects the dynamics within the simulations more generally,
using sample spectra to illustrate the results.  As in the previous
section, we consider three different target accretion rates.

\begin{figure}
\includegraphics[scale=0.33]{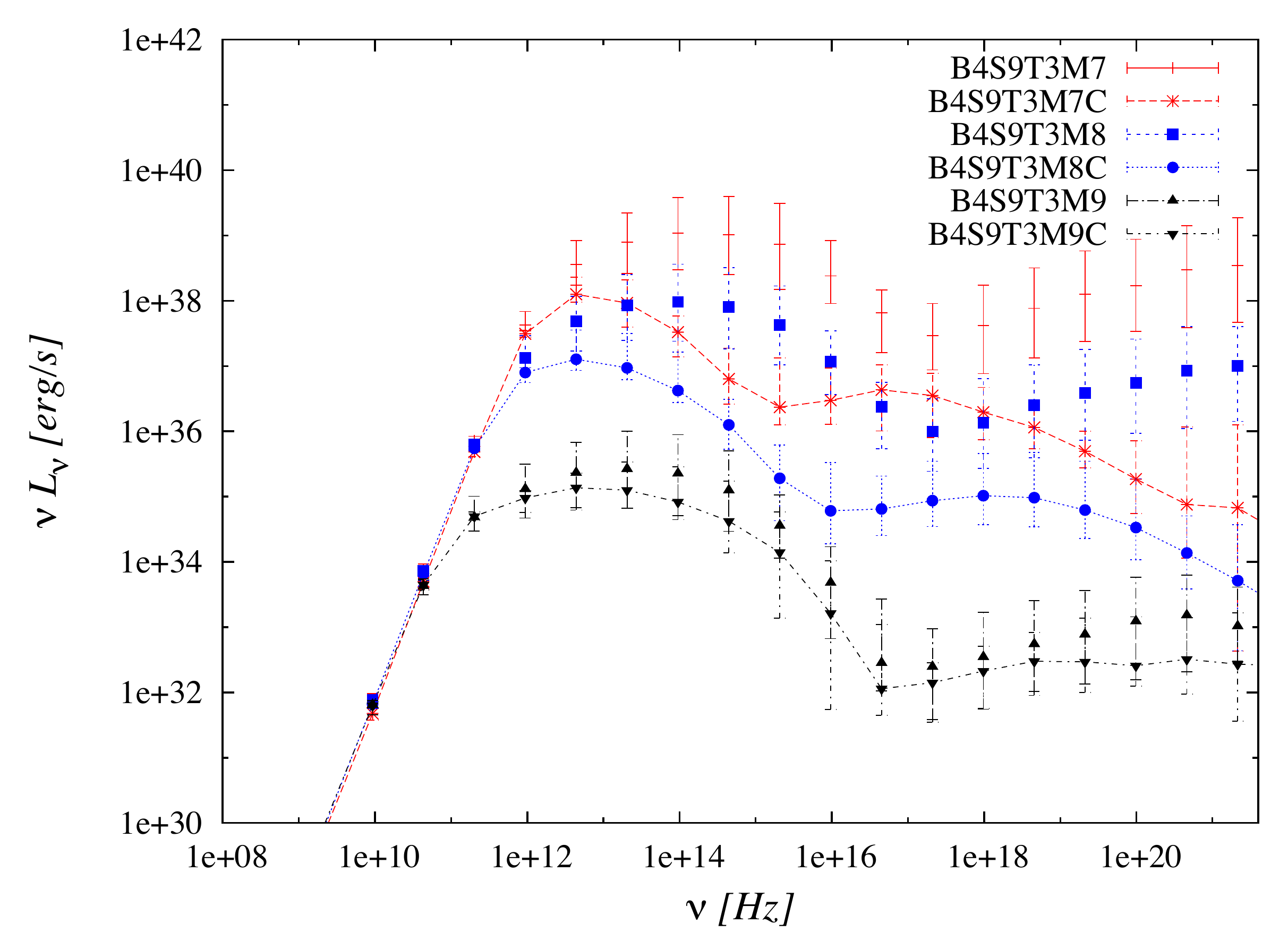}
\caption{\textit{Broadband spectra for simulations B4S9T3M9C,
    B4S9T3M8C, B4S9T3M7C and B4S9 (rescaled to three different accretion rates) with an inclination angle of 85 degrees. The median is taken over the interval $t_{\rm min}-t_{\rm max}$, errors as described for Fig.~\ref{ResolutionTest}.}} 
\label{Spectra}
\end{figure} 

Figure \ref{Spectra} shows the median broadband spectral energy
distribution for four different simulations: B4S9T3M9C, B4S9T3M8C,
B4S9T3M7C, and B4S9.  The simulation B4S9 has been scaled to three
different accretion rates to be consistent with the three simulations
with cooling. The spectra represent the synchrotron and inverse
Compton emission of the system at each frequency. They also include
all special and general relativistic effects via general relativistic
ray tracing using the codes \textsc{geokerr} \citep{dexteragol09} and
\textsc{grtrans} \citep{dexterphd}. We can see that for the low accretion
rates ($\dot{M} \sim 10^{-9} {\rm M_{\odot}/yr} = 10^{-8} \dot{M}_{Edd}$), the spectra of simulations
B4S9T3M9C and B4S9 are nearly indistinguishable, while for the high
accretion rates ($\dot{M} \gtrsim 10^{-8} {\rm M_{\odot}/yr}=10^{-7} \dot{M}_{Edd}$), the
spectra obtained without including the cooling losses self-consistently in the simulation are quite different than the ones where
they are included. Thus, in the regime $\dot{M} \gtrsim 10^{-7} {\rm
  \dot{M}_{Edd}}$, we conclude that simulations which do not take cooling losses into
account are likely not providing a realistic representation
of the physics occurring during the accreting process.  In this case, fitting for parameters such as spin, temperature ratio, magnetic filed configuration, inclination, and accretion rate is likely to yield incorrect results.  This would seem to apply to all known LLAGN except perhaps Sgr A*.

\begin{figure}
\includegraphics[scale=0.33]{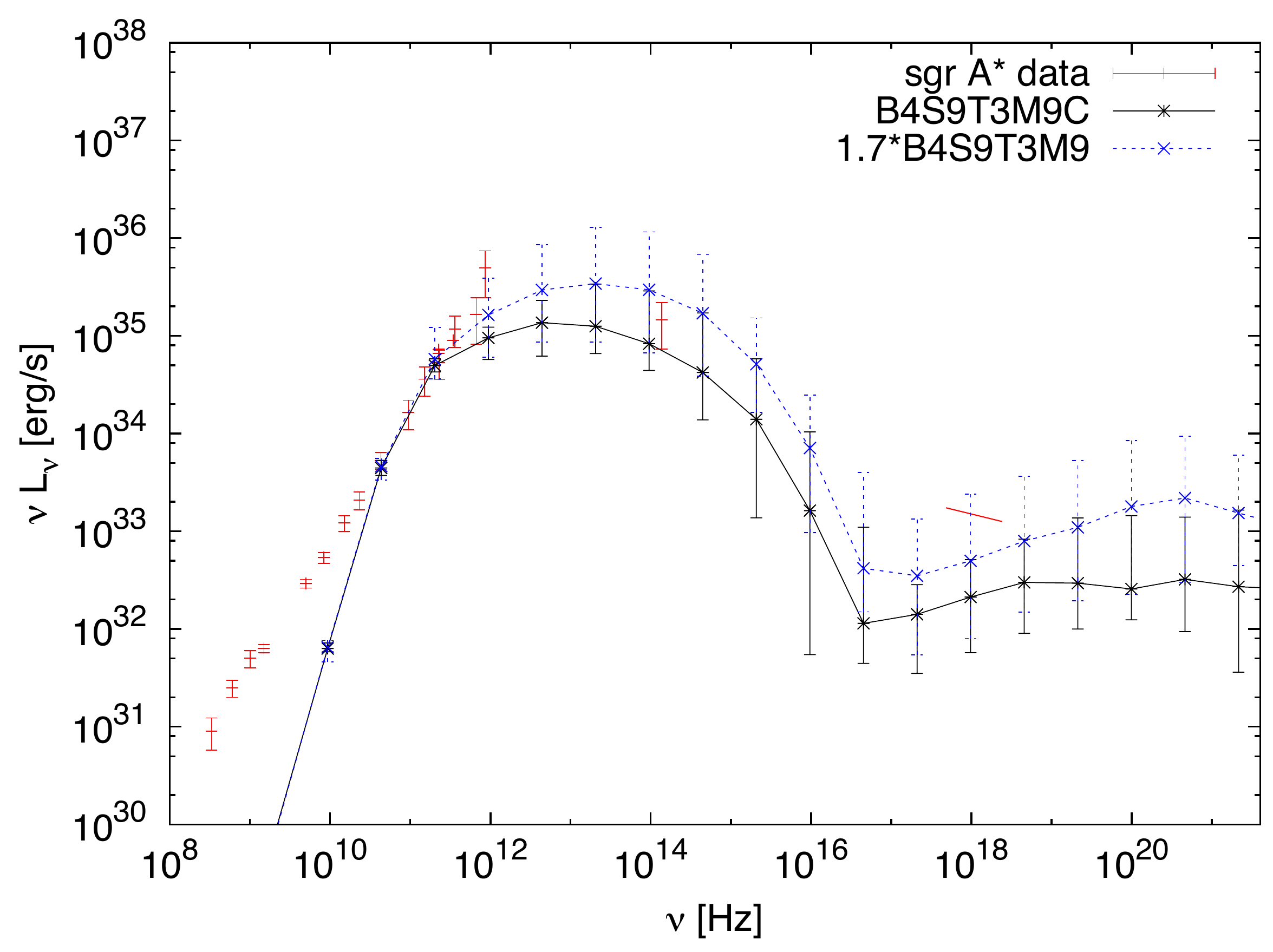}
\caption{\textit{Broadband spectra for simulations B4S9T3M9C and B4S9
    (rescaled to fit the point 230 GHz, 3 Jy) with an inclination
    angle of 85 degrees. The median is taken over the interval $t_{\rm
      min}-t_{\rm max}$.  Observational data for Sgr A* are
    represented in red. \citep{falcke00}.}} 
\label{Fit}
\end{figure}

Of the different simulations considered in this work, the spectra of simulation B4S9T3M9C is the most compatible with the observational data of Sgr A* (see Figure \ref{Fit}). Note
that we are not trying to fit the lowest energy data, as the radio
emission is believed to be emitted from a region much further out than the radial extent of our simulation domain.  In contrast, the so-called ``sub-mm bump'' (see, e.g., \citealt{melia01}) is expected to originate from quite close to the black hole, i.e. the region represented by our simulation.   

To provide concrete values allowing comparison with other works, we
find that we can fit the Sgr A* data at 230 GHz, which has a value of
3 Jy, or $\nu L_\nu \sim 4\times 10^{35}$ erg/s, with an accretion
rate of $2.60 \pm 1.54\times 10^{-9} {\rm M}_{\odot}$/yr via the
non-scalable, cooled simulation B4S9T3M9C (see Table 1). To fit this
data with the corresponding non-cooling simulation B4S9, we must scale
the simulation to have an accretion rate of $2.95 \pm 0.87\times
10^{-9} {\rm M_{\odot}/yr}$ at the BH horizon.  These values are statistically consistent, confirming our earlier point that, at the currently best fit accretion rate, it is not necessary to consider radiative cooling processes when simulating accretion onto Sgr A*.  For a complete parameter study and the best fits to Sgr A* with our models, see D12.

\section{Results: parameter study}

Because we have considered a fairly large parameter space in our study,
we briefly report how our simulations are affected by each of these.  Many of the conclusions reached in this section confirm results of earlier studies by other authors.  We include them
again here (with references to the earlier works) for completeness and to allow comparison.

\subsection{Influence of the initial magnetic field configuration}

As noted in previous studies \citep[e.g.][]{beckwith08, mckinney09},
the initial magnetic field configuration has a major influence on the
launching and continued powering of jets in numerical simulations.
Although our study is more focused on the disc than the jets, we did
make a measure of the instantaneous energy carried in the fluid
and magnetic components of the
jets (defined as material with $b^2/\rho > 1$) at the time of each
data writeout from our simulations.  The fluid and magnetic energy of the jet
are defined as
\begin{equation}
E_\mathrm{FL,jet}(t)=\int \left[ \sqrt{-g}(\rho h u^0 u_0 + P) \right] dr d\theta d\phi
\end{equation}
and
\begin{equation}
E_\mathrm{EM,jet}(t)=\int \sqrt{-g}\left(2P_B u^0 u_0 + P_B - b^0 b_0\right) dr d\theta d\phi ~.
\end{equation}
Table 2 shows these quantities,
averaged over the usual interval ($t_{\rm min}-t_{\rm max}$), for a
subset of our simulations.
We also report the same measure of jet efficiency as \citet{tchekhovskoy11}, namely $\eta = (\dot{M} - \dot{E})/\langle\dot{M}\rangle$, where 
\begin{equation}
\dot{M} (t) = -\int \sqrt{-g} \rho u^r d\theta d\phi
\label{eqn:massFlux}
\end{equation}
is the mass accretion rate and
\begin{equation}
\dot{E} (t) = \int \sqrt{-g} T_t^r d\theta d\phi~
\label{eqn:energyFlux}
\end{equation}
is the energy flux, both taken at the black hole event horizon \(r_\mathrm{BH}\).  Angle brackets indicate a time-averaged quantity.  The negative sign in equation (\ref{eqn:massFlux}) indicates that the mass flux is positive when rest mass is transported into the black hole.  Similarly, $\dot{E}$ is constructed such that a positive value indicates a net flux of energy into the black hole.

\begin{table}
\centering
  \caption{\textit{Jet energies and efficiencies}}
  \begin{tabular}{@{}lccc@{}}
  \hline
                         &  $E_\mathrm{FL,jet}$ &  $E_\mathrm{EM,jet}$ & $\eta$ \\   
   Simulation   &  (erg)     &  (erg)       & \\   
 \hline
B4S9rT3M9C & $3.43\times 10^{38}$ & $6.23\times 10^{38}$ & 0.225 \\
B4S0T3M9C & $1.40\times 10^{39}$ & $1.88\times 10^{39}$ & 0.0625\\
B4S5T3M9C & $7.23\times 10^{39}$ & $9.80\times 10^{39}$ & 0.120 \\
B4S7T3M9C  & $5.50\times 10^{39}$ & $1.07\times 10^{40}$ & 0.161 \\
B4S9T3M9C & $1.03\times 10^{40}$ & $2.19\times 10^{40}$  & 0.464 \\
B1S9T3M9C & $1.07\times 10^{41}$ & $6.47\times 10^{41}$ & 0.559 \\
B4S98T3M9C & $1.72\times 10^{40}$ & $7.60\times 10^{41}$ & 0.525 \\
\hline
\end{tabular}
\end{table}

In our models we start with a relatively weak ($\leq 17.5 {\rm G}$)
magnetic field contained entirely within the disc, initially in the
form of poloidal loops.  We tested two different configurations: a
single set of poloidal loops centered on the pressure maximum of the
torus and following contours of pressure/density (an example is
simulation B1S9T3M9C) and four sets of poloidal loops spaced radially,
with alternating field directions in each successive loop (an example
is simulation B4S9T3M9C).

The 1-loop configuration yields a significantly more energetic outflow (mostly
associated with the ``jets'') -- more than an order of magnitude
stronger than for the more complex, though perhaps more realistic,
configuration consisting of 4 initial loops.  We have confirmed that, as expected, the overall power
of both components scales roughly linearly with the mass accretion
rate, meaning the efficiencies of the outflows are not strongly
effected by the cooling.

\begin{figure}
\includegraphics[scale=0.14]{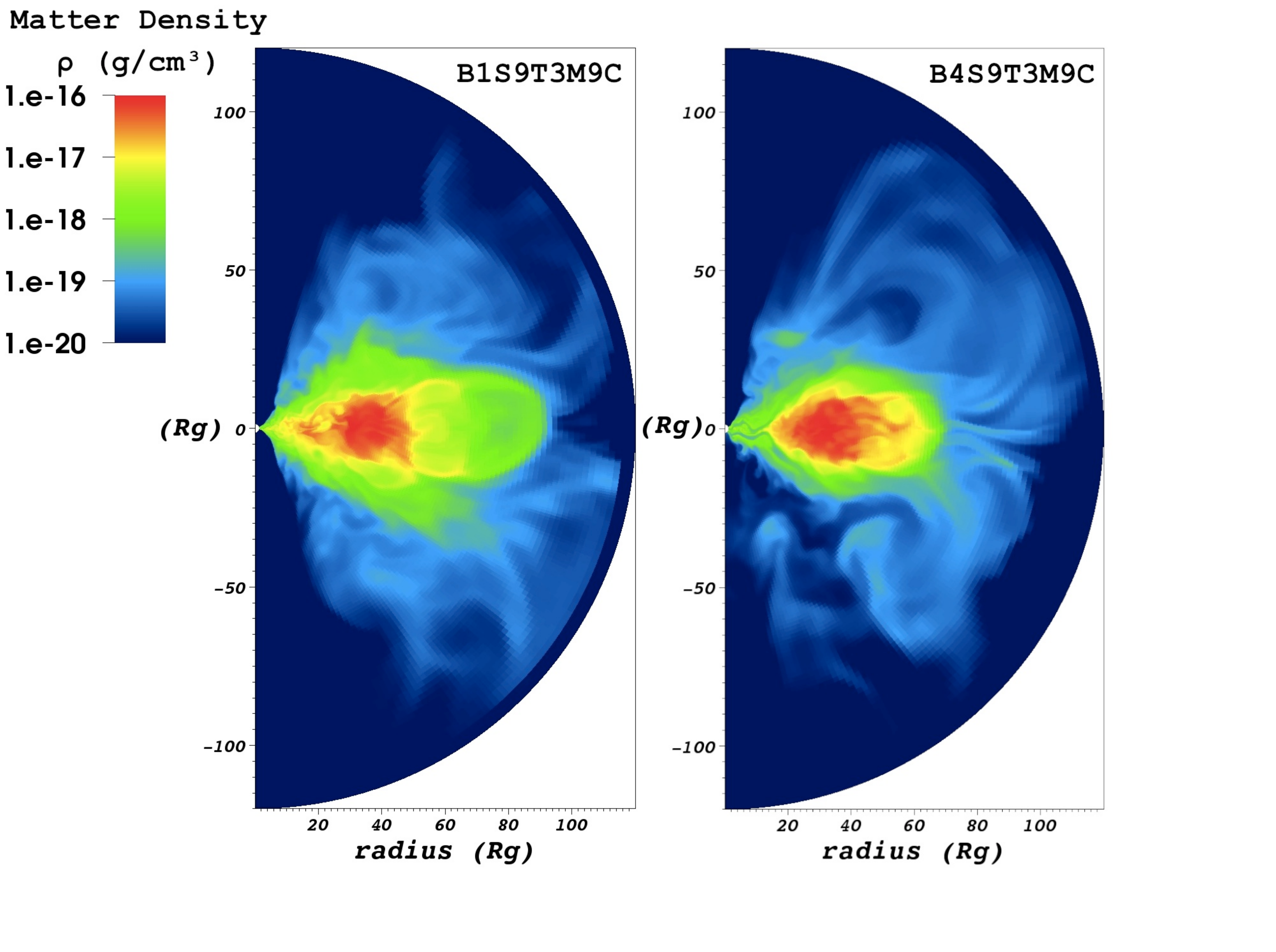}
\caption{\textit{Density (in $g/cm^{3}$) of the system for simulations
    B1S9T3M9C and B4S9T3M9C at the same time step (t= 3 orbits). The two simulations started with the same initial torus configuration as shown in figure \ref{InitialConfiguration}. The only initial difference between the two models is that the left panel started with a single magnetic loop configuration (as shown in the bottom panel of figure \ref{Initial_MagneticField}), while the right panel started with a 4 magnetic loop configuration  (as shown in the top panel of figure \ref{Initial_MagneticField}). We
    can see that in the case of the single initial magnetic loop, the channel solution extends through the entire disc,
    resulting in a higher mass accretion rate at the black hole horizon and a more extended disk.}} 
\label{Rho_1B4B}
\end{figure}

Aside from the outflow, the initial magnetic field configuration also
has a strong influence on the accretion rate as seen by comparing the
measured $\dot{M}$ in Table 1 for simulations B4S9T3M9C and B1S9T3M9C.
The 1-loop simulation has a substantially higher average $\dot{M}$, as
well as much larger statistical fluctuations.  This difference is likely a consequence of the axisymmetric nature of these
simulations.  In our simulations, accretion is driven by the MRI.  For
the 1-loop case, the dominant channel solution is able to disrupt
nearly the entire disc, driving large amounts of material onto the
black hole in multiple short, intense episodes.  In the 4-loop cases,
the channel solutions are restricted in their radial range and can not
disrupt the disc as effectively; accretion proceeds more smoothly, and
at an overall lower normalization. [The channel solution \citep{hawley92} is a particularly violent form of the poloidal MRI and is characterized by prominent radially extended features.] These effects are illustrated in Figure \ref{Rho_1B4B}.

\begin{figure}
\includegraphics[scale=0.35]{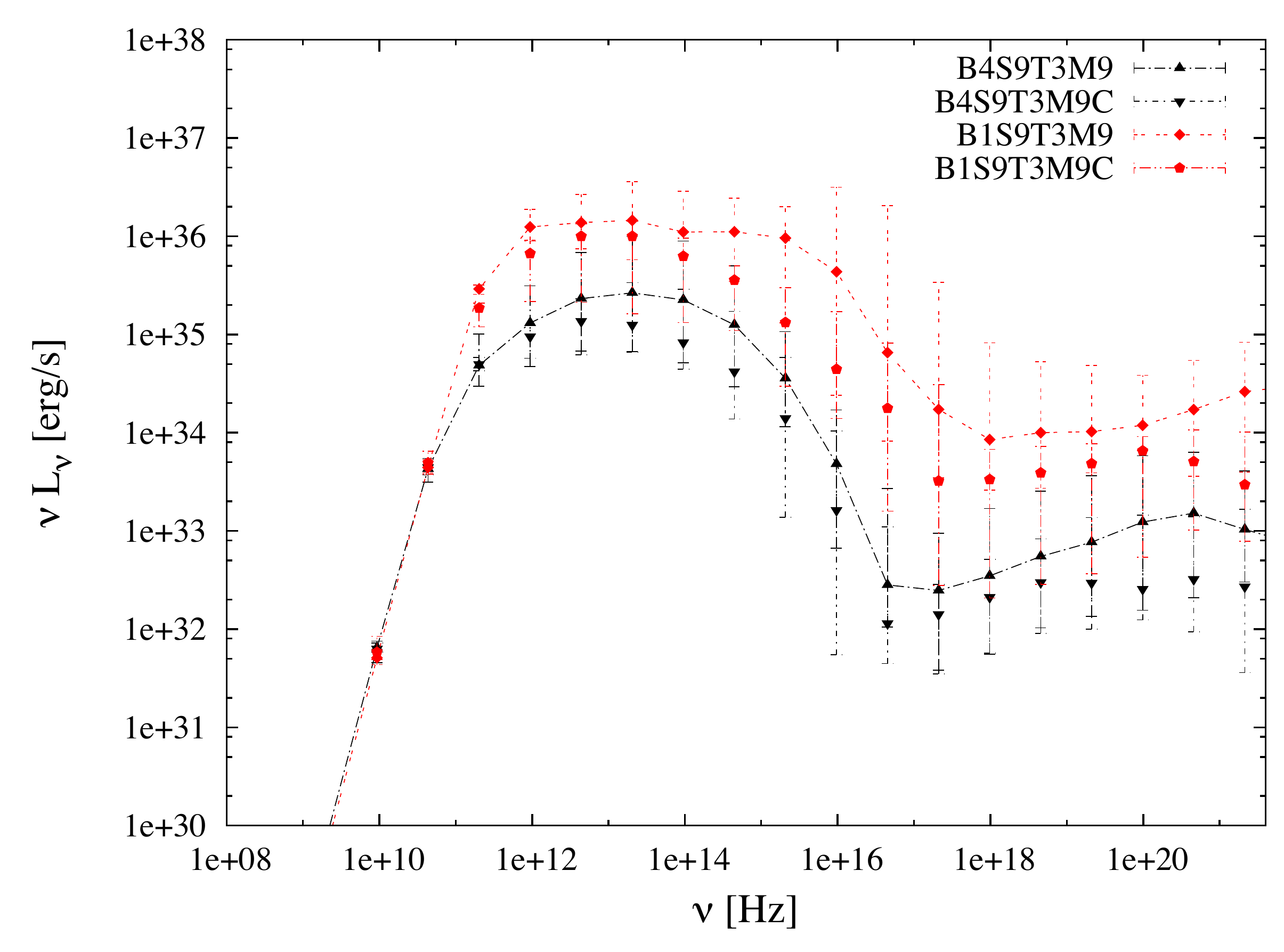}
\caption{\textit{Broadband spectra for simulations B4S9T3M9C,
    B4S9T3M9, B1S9T3M9C, and B1S9T3M9. The median is taken over the
    interval $t_{\rm min}-t_{\rm max}$.}}
\label{SED_B1B4_RT_median}
\end{figure}

Figure \ref{SED_B1B4_RT_median} 
compares the spectra for the two different initial magnetic field configurations, the
``1-loop'' configuration (in black), and the ``4-loop'' configuration
(in red) .  It is clear that the magnetic field configuration has a
strong influence on the resulting spectra and is therefore one of the
big uncertainties affecting all such models of the accretion flow.
The lack of \textit{a priori} knowledge of the field configuration
introduces an effective error in all simulation conclusions that is
comparable to the difference introduced by including cooling.  Thus,
while we can treat the cooling losses self-consistently, it is
difficult to overcome the variations that different choices in initial
magnetic field configurations will introduce.  

\subsection{Influence of the Spin}

Via its formation and accretion history, Sgr A* is likely a rotating
SMBH, but as with all black holes it is currently difficult to
reliably determine its spin. Although the observed spectrum should
depend on spin, it also depends on the mass accretion rate, which is somewhat constrained, as well as the geometry of
the accretion flow.  Nevertheless, spectral fitting of Sgr A* has
already placed some meaningful constraints on the value of $a_*$
\citep{broderick11,shcherbakov10}, although there can be additional
complicating factors, such as disk tilt \citep{dexterfragile11,dexterfragile12}.
Other methods of inferring the ISCO, and then the spin, such as
observing relativistically-broadened line profiles or modeling the
thermal emission from the disc, require the presence of a
geometrically thin, optically thick disc, which is not observed at the
low accretion rates of Sgr A*.  The ultimate determination of spin
will likely require future observations from mm-VLBI
\citep{doeleman08,fish11, broderick11}.

\begin{figure}
\includegraphics[scale=0.35]{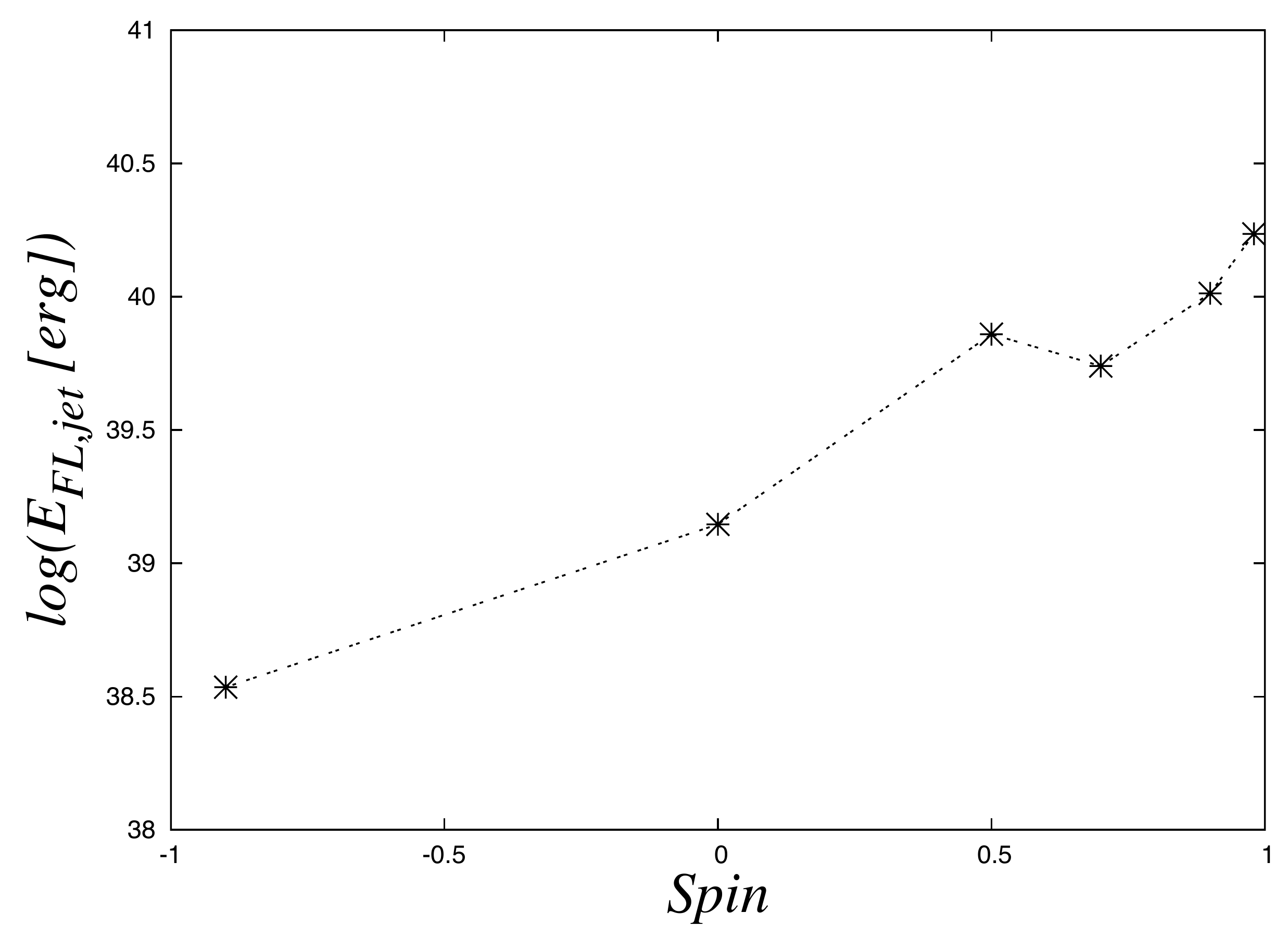}
\caption{\textit{ Jet fluid energy as a function of spin. The six data points represent the averaged jet energies for B4S9rT3M9C, B4S0T3M9C, B4S5T3M9C, B4S7T3M9C, B4S9T3M9C, and B4S98T3M9C.}} 
\label{JetKineticEnergy_spin}
\end{figure} 

\begin{figure}
\includegraphics[scale=0.35]{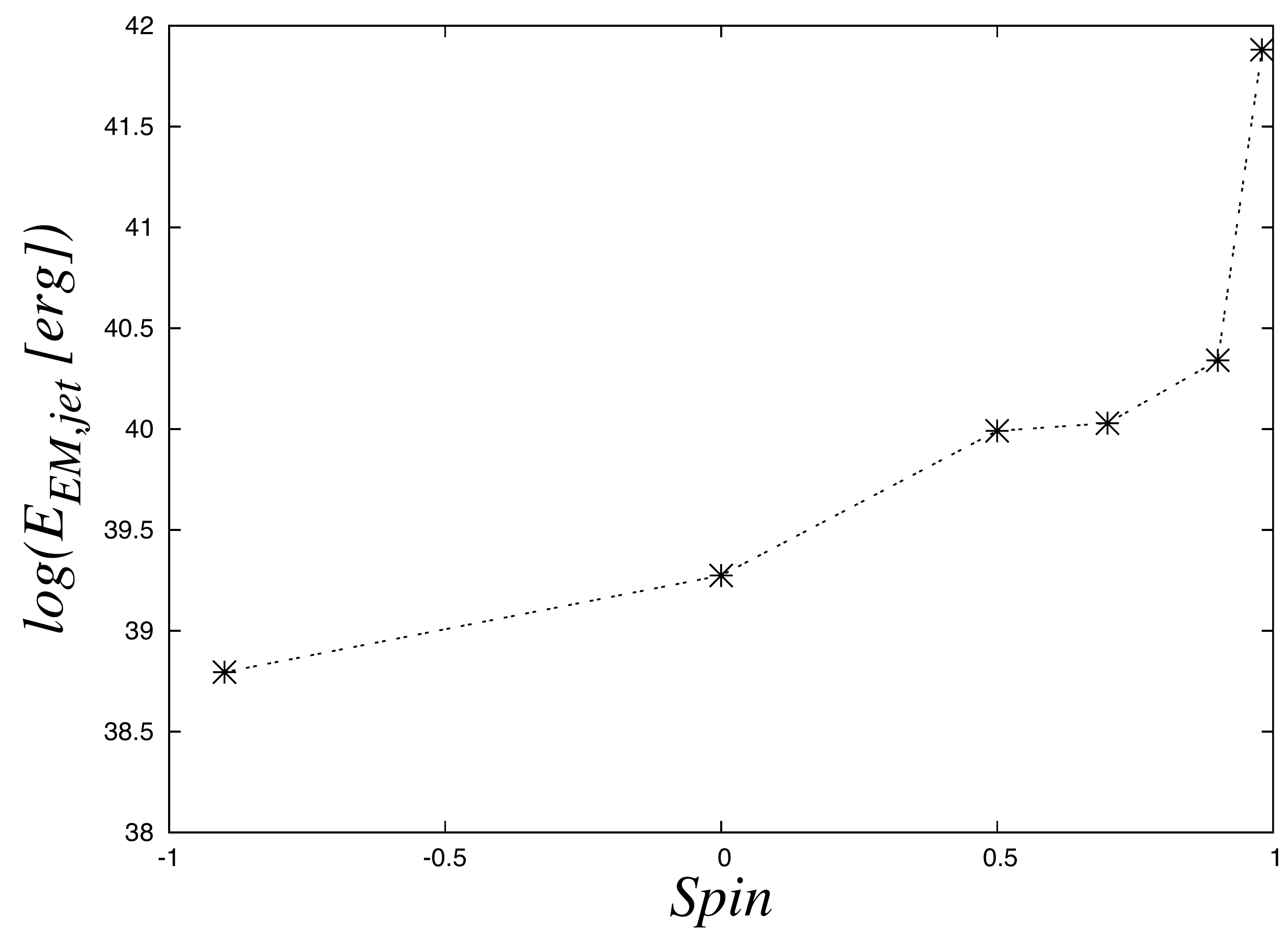}
\caption{\textit{ Jet magnetic  energy as a function of spin. The six data points represent the averaged jet energies for B4S9rT3M9C, B4S0T3M9C, B4S5T3M9C, B4S7T3M9C, B4S9T3M9C, and B4S98T3M9C.}} 
\label{JetMagneticEnergy_spin}
\end{figure}

We investigated the spin dependence of our results by performing
simulations at six different values of $a_*$: -0.9, 0, 0.5, 0.75, 0.9,
and 0.98, where the minus sign indicates retrograde rotation (disc and
black hole rotating in opposite senses).  We will discuss the effects
of spin on our spectral fits in companion paper D12.  Concerning the disc evolution, there does not seem to be any clear trend in 
$\dot{M}$, see Table 1.  If anything there appears to be a peak
in the distributions, with the highest mass accretion rates occurring
at intermediate spins, $0.3 \lesssim a_* \lesssim 0.7$.  The outflow
power on the other hand, shown in Table 2, reveals a clear trend with spin, as
expected based on theoretical grounds
\citep[e.g.][]{blandford77,tchekhovskoy10} and from previous numerical
work \citep{devilliers05,hawley06,mckinney06,tchekhovskoy12}.  This trend is seen in
both the fluid and magnetic components of the outflow and the
correlations are shown in Figure \ref{JetKineticEnergy_spin}.

The effect of spin on jet power is a question that has been debated
recently in the literature, specifically using claimed spin
observations from black hole X-ray binaries.  So far the observational
results are not definitive (see, e.g., the opposing conclusions found
in \citealt{fender10} and \citealt{narayan12}).  Our results, while
focusing on SMBHs, would be predicted to scale to smaller mass black
hole systems via the Fundamental Plane of BH accretion relation
\citep{merloni03,falcke04}.  Semi-analytical models for the compact
jets observed in weakly accreting systems predict a dependance of
radio luminosity on total jet power of $\sim Q_j^{1.4}$
\citep[e.g.,][]{falcke95}.  Thus if we consider the total power as the
sum of the fluid and magnetic components in
Figure~\ref{JetMagneticEnergy_spin}, we would also predict a correlation
between observed radiative power and spin.

\subsection{Disk Thickness}

In all our simulations we started with the same disk thickness ($\Delta r / r = $constante $\simeq 1$) in order to start with a thick disk, appropriate for the Galactic center. Then the shape of the disk is modified during the simulation and is especially sensitive to the cooling as shown in figure \ref{Rho_Cool}.  We also investigated thinner disk simulation, the thinner disk emmit a little bit less that the thick disk but they could be compatible within one sigma.

\section{Conclusions}

For the first time we have been able to assess the importance of the
radiative losses in numerical simulations of LLAGN, specifically Sgr
A* [see also \citet{moscibrodzka11}]. We show that radiative losses can
affect the dynamics of the system and that their importance increases
with accretion rate.  We set a rough limit of $\dot{M} \gtrsim
10^{-7}\dot{M}_{\rm Edd}$, above which radiative cooling losses should be
included self-consistently in numerical simulations.  Otherwise, many
important derived dynamical quantities, such as density, magnetic
field magnitude, and temperature, may be off by an order of magnitude
or more, especially when the accretion rate reaches $\dot{M} \simeq
10^{-6}\dot{M}_{\rm Edd} $, which correspond to $10^{-7} M_{\odot}~{\rm
  yr^{-1}}$ for Sgr A*.  Since several recent works suggest that
accretion physics is similar across the mass scale, this accretion
rate in Eddington units should likely be an important limit for all
black holes.  Thus, we predict that the inclusion of self-consistent
radiative cooling above $\sim10^{-6} M_{\rm Edd}$ should be important for
LLAGN in general.

Overall, this study allows us to have a more consistent model of
accretion, even for the well-studied and under-luminous source, Sgr
A*.  Not only do we have a more realistic model by including the
physics of radiative losses, but we are also able to show the
influence of the accretion rate on the resulting spectra.

The spin of the central black hole and the initial magnetic field
configuration of the torus also have important consequences on the
dynamics of the system and the resulting spectra. By including the
cooling losses in our study, we can discuss the influence of these
free parameters with more accuracy.  For instance, initial magnetic
field configurations consisting of a single set of poloidal loops
result in significantly more powerful outflows than the four-loop cases,
and we find that the jet power increases with the spin of the
central black hole.

As mentioned in Section \ref{sec:time_interval}, there is concern as
to whether or not our simulations have run long enough to reach
meaningful equilibrium states.  This concern is especially pertinent for our
case using 2.5D simulations, as these can never truly reach a steady state.  The
reason is that, after a period of initial vigorous growth, the MRI in
2.5D simulations steadily decays because the dynamo action that normally
sustains it requires access to non-axisymmetric modes that are
obviously inaccessible.  We, therefore, chose a time interval for
analysis when the mass accretion rate closely approximated our target
value for most simulations.  Clearly simulations in 3D with a longer
duration (to ensure a proper equilibrium is reached) would be
beneficial, for comparison.

One other concern with only performing 2.5D, axisymmetric simulations
is that it precludes the possibility of exploring the effects of
misalignment between the angular momentum of the gas and the black
hole.  In reality, most supermassive black holes are unlikely to be
accreting from matter sharing the same orbital plane as the black hole
spin.  \citet{dexterfragile12} recently showed that accounting for
such a ``tilted'' accretion flow can dramatically alter the best-fit
characteristics of Sgr A* and produce important new features in the
spectra.

As a final note, one can justify neglecting full radiative transfer in
simulations of Sgr A* and likely most other weakly accreting LLAGN
because the inner regions are generally thought to be optically thin.
However, to treat the outer regions of the accretion flow, or higher
luminosity sources, a more thorough treatment of radiative transfer
will need to be implemented into the simulations.  Similarly, to
approach important questions such as the mass loading and particle
acceleration in the jets, likely resistive (non-ideal) MHD will need
to be considered.  We see this work as an important step towards these
next technological ``horizons''.

\section*{Acknowledgments}
We acknowledge support from The European Community's Seventh Framework
Programme (FP7/2007-2013) under grant agreement number ITN 215212
Black Hole Universe.  SM and SDi gratefully acknowledge support from a
Netherlands Organization for Scientific Research (NWO) Vidi
Fellowship.   This work was also partially supported by the National Science
Foundation under grants AST 0807385 and PHY11-25915 and through TeraGrid resources
provided by the Texas Advanced Computing Center (TACC).  We thank SARA
Computing and Networking Services (www.sara.nl) for their support in
using the Lisa Computational Cluster.  PCF acknowledges support of a
High-Performance Computing grant from Oak Ridge Associated
Universities/Oak Ridge National Laboratory.

\label{lastpage}
\end{document}